\documentclass{aa}  
\usepackage[utf8]{inputenc}
\usepackage{hyperref}
\hypersetup{
     colorlinks = true,
     linkcolor = blue,
     anchorcolor = blue,
     citecolor = blue,
     filecolor = blue,
     urlcolor = blue
     }
\usepackage{graphicx}
\usepackage{txfonts}
\usepackage{ragged2e} 
\usepackage{xcolor}
\usepackage{textcomp}

\definecolor{darkred}{RGB}{139,0,0} 

\newcommand*{\hi}{H\,{\sc i}\xspace}

\makeatletter
\newcommand{\Figref}[1]{\@ifundefined{r@#1}{Fig.}{Fig.~\ref{#1}}}
\makeatother
\begin{document}

   \title{Measuring radial flows in disc galaxies}

   \author{Yi He
          \inst{1}, 
          Filippo Fraternali\inst{1}, Pavel E. Mancera Piña\inst{2}, and Enrico M. Di Teodoro\inst{3,4}
          }

   \institute{Kapteyn Astronomical Institute, University of Groningen, Landleven 12, 9747 AD, Groningen, The Netherlands\\
              \email{linn@astro.rug.nl}
         \and Leiden Observatory, Leiden University, P.O. Box 9513, 2300 RA Leiden, The Netherlands
 \and
    Universit\`{a} di Firenze, Dipartimento di Fisica e Astronomia,  Largo E. Fermi 2, 50125 Firenze, Italy
    \and
    INAF - Osservatorio Astrofisico di Arcetri, Largo E. Fermi 5, I-50125 Firenze, Italy
             }

   \date{Received month day, xxxx; accepted month day, xxxx}

 
  \abstract
   {Numerous theoretical studies  suggest that gas accretion from the intergalactic medium is essential to feed star formation in galaxies throughout cosmic time. However, the mechanism of gas accretion remains poorly understood, as direct evidence is lacking. In some models, gas accretion is expected to take place in the outer discs of galaxies, from where it should be transferred to the inner star-forming discs through radial flows. Finding such flows and quantifying them, therefore, opens the possibility of inferring the detailed properties of the accreting gas. Unfortunately, detecting radial flows is challenging due to the large disparity between radial and rotational velocities. Additionally, the distortions in velocity fields produced by these flows resemble the effects of warped structures, further complicating their quantification. To address this, we have developed a new methodology to measure axisymmetric radial velocities in disc galaxies. Relying on the 3D kinematic fitting software \texttt{$^{\rm 3D}$Barolo}, we have designed a dedicated workflow to measure axisymmetric radial motions in disc galaxies and to obtain robust uncertainty estimates through a bootstrapping approach. We validated this approach using realistic axisymmetric mock galaxies as well as a small sample of `observed' Milky Way-like galaxies drawn from the TNG50 hydrodynamic simulation. These tests demonstrate that our method works well and is capable of overcoming the degeneracy between radial motions and  warps.
   The TNG50 galaxies we analysed have large regions in their discs with non-axisymmetric inflows and outflows, but they show, on average, no global trend of coherent radial flows.
   }

    
   \maketitle

\section{Introduction}
Gas accretion is a fundamental driver of galaxy evolution. 
The large population of star-forming disc galaxies in the local Universe has been forming stars over extended periods of time. 
However, their available cold gas reservoirs are insufficient to sustain star formation over such long durations: typical gas depletion timescales are only a few billion years, which are significantly shorter than the inferred star-forming lifetimes of these systems \citep[e.g.][]{2000MNRAS.318..658B,2008AJ....136.2782L,2012MNRAS.426.2166F,2013ApJ...778....2S}.
This mismatch indicates that an additional and continuous supply of gas is required to maintain ongoing star formation in disc galaxies.
Further independent evidence for ongoing gas accretion also arises from chemical evolution considerations. Accreted gas is expected to be metal-poor, and the long-standing need for low-metallicity infall has been recognised in chemical evolution models of the Milky Way for several decades \citep[e.g.][]{1991MNRAS.251...84P,1996MNRAS.280..720V,2012ceg..book.....M,2015A&A...580A..17G}.

Galaxies can acquire fresh gas through both galaxy mergers and smooth gas accretion from their surroundings. However, observational evidence suggests that, at least in the local Universe, mergers alone cannot provide a sufficient gas supply to sustain star formation in spiral galaxies \citep{2008A&ARv..15..189S,2014A&A...567A..68D,2016A&A...590A..51L}. 
Estimates of the neutral gas accretion rate based on high-velocity clouds indicate values that are a factor of 3–5 lower than the observed star formation rates in Milky Way and M31-type galaxies \citep{2006ApJ...645.1164P,2012ApJ...750..165R}. 
Additionally, deep 21-cm observations from the MeerKAT H\,{\sc i}
Observations of Nearby Galactic Objects: Observing Southern Emitters
(MHONGOOSE; \citealt{2024A&A...688A.109D}) and the FAST Extended Atlas
of Selected Targets Survey (FEASTS; \citealt{2023ApJ...944..102W}) reveal a surprising `\hi desert' \citep{Fraternali2024} in the inner circumgalactic medium (CGM; $\lesssim 100$ kpc), with a lack of cool \hi clouds down to $N_{\text{\hi}} \sim 10^{17} \text{ cm}^{-2}$ \citep{2024A&A...688A.109D, 2025A&A...697A..86M}.
This confirms earlier results on a few isolated galaxies observed with other interferometers at a high sensitivity \citep[e.g.][]{Fraternali+2002, Oosterloo+2007, Kamphuis+2022}.
Overall, direct observational evidence for gas accretion remains scarce, with highly uncertain accretion rates \citep{2008A&ARv..15..189S,2012ARA&A..50..491P}.
How gas accretion into galaxies takes place to sustain star formation over long timescales remains a key open question.

The current theory of gas accretion allows for two modes, cold and
hot, depending on galaxy mass and redshift \citep[e.g.][]{2003MNRAS.345..349B,2004MNRAS.347.1093B}.
In cold-mode accretion, which is typical of low-mass galaxies and galaxies at high redshifts, cold gas streams from intergalactic filaments penetrate the halo and reach the disc directly, typically depositing material at large galactocentric radii \citep{2005MNRAS.363....2K}. 
In hot-mode accretion, the gas falling into the galactic potential well is instead shock-heated to a temperature close to the virial temperature and forms a large atmosphere surrounding the galaxy \citep{2005MNRAS.363....2K, Gronnow+2024}.
Accretion onto the disc can then take place via the cooling of this gas, either spontaneously \citep{2020MNRAS.492.6042S} or triggered by galactic fountains \citep{Fraternali2017}.
In some cases, filaments could also manage to penetrate the hot atmosphere \citep{Mandelker+2020}, although this becomes unlikely at lower redshifts \citep{2023MNRAS.524.2351A}.

Due to the scarce observational evidence for gas accretion, cold- and hot-mode accretion scenarios remain largely untested.
However, these scenarios can in principle be distinguished, as they predict distinct properties for the gas accreting onto galaxy discs.
In general terms, one can expect that in hot-mode accretion, gas cools out of a virial-temperature hot circumgalactic medium and accretes over the disc, where it can directly feed star formation. 
Conversely, in cold-mode accretion, gas joins the disc preferentially in the outer parts, from which it should be transported inwards to feed inner star formation.
Fountain accretion predicts a somewhat intermediate situation \citep{Marasco+2012}.
Clearly, in these different scenarios, galaxy discs experience a different amount of gas transport from outer to inner regions \cite[e.g.][]{Pezzulli+2016,2019ApJ...875...54H, 2022MNRAS.509.4149T}, which can, in principle, be measured in observations in terms of radial flows.

Early studies, based on 2D kinematic maps, reached contradicting conclusions. \citet{2004ApJ...605..183W} combined CO and \hi velocity fields for a sample of seven nearby spiral galaxies and decomposed the line-of-sight velocity into Fourier components to infer radial motions, but they found no clear evidence for a systematic radial inflow. 
Instead, they argued that bars, spiral arms, and warps can mask or mimic radial flow signatures. 
Using data from The \hi Nearby Galaxy Survey
(THINGS; \citealt{2008AJ....136.2563W}) for NGC 2976, \cite{2010MNRAS.404.1733S} identified non-circular streaming motions in the strongly barred galaxy. 
\cite{2016MNRAS.457.2642S} analysed \hi data from the THINGS survey for ten moderately inclined nearby galaxies and found evidence for radial gas flows in approximately half of their sample, with inferred mass inflow rates comparable to or exceeding the star formation rates. 
\cite{2022MNRAS.509..100B} performed a kinematic analysis of H$\alpha$ velocity fields incorporating bisymmetric flows and found one galaxy exhibiting radial velocities as large as $50\ \mathrm{km\ s^{-1}}$, while the remaining galaxies showed complex non-circular motions in their inner discs, including radial and bisymmetric components.

Given its extent, which is significantly beyond the optical radius, \hi is the ideal tracer of radial flows in the outer discs, where contamination from streaming motion in spiral arms and bar-induced flows is minimised.  
\citet{2021ApJ...923..220D} recently introduced a three-step 3D modelling approach, in which the full \hi data cube was fitted directly rather than reduced to a 2D velocity field, to extract radial gas flows in 54 nearby disc galaxies. In their method, the disc geometry (position angle and inclination) and rotation curve were first determined by fitting a tilted-ring model with increased weighting towards the kinematic major axis, where rotational motions dominate. In a subsequent step, the radial velocity component was fitted while keeping the geometry fixed and assigning a higher weight to regions near the kinematic minor axis, where the radial flow signature is expected to be strongest. 
Their analysis reported that the majority of galaxies exhibit only weak inflow or outflow motions, with amplitudes of a few $\mathrm{km\ s^{-1}}$. 

While \citet{2021ApJ...923..220D}'s approach introduces significant improvements with respect to previous attempts, including the fitting of the entire emission-line datacube, it implicitly assumes that the distortions observed near the kinematic major axis arise fully from geometric warps, and that the contribution from radial motions is negligible in those regions. 
This assumption was further tested by \cite{2023ApJ...944..143W} who showed that both disc warps and radial flows independently distort the velocity field relative to pure circular rotation, and that their kinematic signatures can be difficult to disentangle, becoming increasingly ambiguous for low-inclination systems. 
Moreover, \cite{2023ApJ...944..143W} showed that warps and radial flows can distort the kinematic major and minor axes either in the same or in opposite directions, such that warps may either enhance or suppress the kinematic distortions produced by radial flows. 
Further support for this degeneracy was provided by \cite{2025ApJ...988..122S}, who examined correlations between radial and transverse velocity components and showed that fitting disc warps prior to allowing for radial motions can artificially enhance inferred warp signatures. 
It is noteworthy that \cite{2016MNRAS.457.2642S} pointed out that fitting the disc geometry and radial velocity simultaneously can mitigate this compensation effect, thereby reducing the degeneracy between warps and radial flows. However, their analysis was based on 2D velocity fields.

In summary, distinguishing between disc warps and radial gas flows remains a challenging problem, as the kinematic signatures of these two phenomena can easily be confused. Since HI warps are ubiquitous among nearby disc galaxies \citep{1990ApJ...352...15B,1994A&A...290L...9R,2002A&A...394..769G}, this degeneracy affects a large fraction of systems and therefore represents a major obstacle to accurately measuring radial gas flows.
Even when 3D modelling techniques are employed, assumptions about the fitting sequence can bias the inferred balance between geometric distortions and genuine radial motions. 
In particular, if the warp geometry, especially the position angle, is fitted first, the kinematic signature of radial flows can be partially absorbed into an artificial geometric warp.
These limitations have hindered our ability to determine whether the gas component in disc galaxies exhibits coherent radial motions, to assess the possible prevalence of inflow, and, if present, to quantify the corresponding inflow rates, with the ultimate goal of gaining insights into the gas accretion mechanisms. 
Recent theoretical work has suggested that outer \hi\ warps may be linked to hot gas accretion from the circumgalactic medium \citep{2025arXiv251103793S}. 
Although exploring the physical origin of warps is beyond the scope of this paper, this possibility reinforces the need to disentangle geometric warps from radial motions when using disc kinematics to constrain the dominant gas accretion mechanisms in galaxies. 

In this paper, we develop a new method that employs the 3D modelling software \texttt{$^{\rm 3D}$Barolo} \citep{2015MNRAS.451.3021D} to simultaneously fit the disc geometry and radial velocity. This strategy allows us to better disentangle the degeneracy between geometric warps and radial gas flows while fully exploiting the information contained in the datacube. 

This paper is organised as follows. In Section~\ref{sec:methods}, we describe our methodology, in which we use $^{\rm 3D}$Barolo to simultaneously fit the disc geometry and radial velocity. 
We further create mock galaxies and synthetic observations drawn from cosmological hydrodynamical simulations, which we use to validate our method. 
In Section~\ref{sec:results}, we present the results obtained from the mock galaxies, and we further show the recovery of radial velocities from the simulated galaxies. 
In Section~\ref{sec:discussion}, we discuss the implications of our results and the predictions of our method. Finally, in Section~\ref{sec:conclusion}, we summarise our main conclusions.

\section{Methods}
\label{sec:methods}

\subsection{Fitting process}
\label{sec:fitting_process}

After extensive tests using a variety of approaches and procedures, we have found an optimal method to measure the radial velocity profile in galaxy discs. Our method adopts a four-step fitting strategy, with a particular focus on disentangling the effects of warps and radial flows. 
It relies on the software \texttt{$^{\rm 3D}$Barolo}, which performs 3D tilted–ring kinematic modelling of emission-line datacubes to recover the kinematic and geometrical properties of rotating disc galaxies. The galaxy disc is represented as a set of concentric tilted rings, each characterised by its own geometrical and kinematical parameters, to recover the kinematic and geometrical properties of spiral galaxies.
The approach we have implemented is described below.

The disc kinematics are extracted through two consecutive runs of the \texttt{TWOSTAGE} option in \texttt{$^{\rm 3D}$Barolo}.
Each \texttt{TWOSTAGE} run consists of an initial ring-by-ring fit followed by a second fit in which selected geometrical parameters are regularised. 
The first \texttt{TWOSTAGE} run, corresponding to Steps~1 and 2, is used to obtain robust rotation velocity $V_{\rm rot}$ and velocity dispersion $\sigma$ radial profiles. 
The second \texttt{TWOSTAGE} run, corresponding to Steps~3 and 4, is then used to measure the radial velocity while allowing for smooth PA variations. 
The procedure is summarised as follows.

\begin{itemize}

\item Step 1.
We first obtain an initial ring-by-ring fit of the disc kinematics.
In this fit, the rotation velocity $V_{\rm rot}$, velocity dispersion $\sigma$, systemic velocity $V_{\rm sys}$, inclination $i$, galaxy centre, position angle (PA), and radial velocity $V_{\rm rad}$ are treated as free parameters and are fitted simultaneously for each ring.

This fit adopts the standard $\cos^2\theta$ weighting scheme in \texttt{$^{\rm 3D}$Barolo} ($\texttt{WFUNC}=2$), where $\theta$ is the azimuthal angle measured from the kinematical major axis. 
This scheme gives higher weight to pixels close to the major axis, where $\theta \simeq 0^\circ$ and the projected rotational signal ($V_{\rm rot}\cos\theta$) is strongest. 
For the mask, we adopt the \texttt{SMOOTH\&SEARCH} method, which spatially smooths the cube by a factor of two before running a 3D source finder to identify the galaxy emission. 
We set a primary signal-to-noise threshold of $\texttt{SNRCUT}=5$ and a secondary growth threshold of $\texttt{GROWTHCUT}=3$ for the source finder. These choices related to masking can be adapted to the specific datacube one works with.

\item Step 2.
We then perform the second fit of the \texttt{TWOSTAGE} run to refine the rotation curve and velocity dispersion profile. 
Before this fit, $V_{\rm sys}$ and the galaxy centre are fixed to constant values based on Step~1. 
The ring-by-ring PA and inclination profiles obtained in Step~1 are regularised to suppress unphysical fluctuations caused by noise, beam smearing, or local non-circular motions. 
Specifically, we regularise the PA profile by fitting it with a smooth Bézier curve as a function of radius, ensuring a gradual and continuous radial variation. 
For the inclination, we either adopt a constant value or model its radial variation with a linear function or a Bézier curve, depending on the level of variation inferred from Step~1. 
The fit is then performed with these regularised geometrical profiles, while only $V_{\rm rot}$, $\sigma$, and $V_{\rm rad}$ are left free. 
From this step, we only adopt the resulting $V_{\rm rot}$ and $\sigma$ profiles, which are fixed in the subsequent radial velocity fit; $V_{\rm rad}$ is left free here only to reduce the risk that non-circular motions are absorbed into the rotation curve or disc geometry.

\item Step 3.
We then proceed to the main radial flow fit. 
In this step, $V_{\rm rot}$, $\sigma$, and $i$ are fixed to the values obtained in Step~2, while the PA and $V_{\rm rad}$ are left free. 
This is the main difference between our fitting approach and the procedure adopted by \cite{2021ApJ...923..220D}, who fit the PA first, and keep it fixed while fitting $V_{\rm rad}$. 
We use the $|\sin\theta|$ weighting scheme ($\texttt{WFUNC}=-1$), which preferentially weights the minor-axis regions and is therefore better suited to constrain radial motions, and we discuss further weighting schemes in Sect.~\ref{sec:weighting scheme}.

\item Step 4.
Finally, the PA profile obtained in Step~3 is regularised using a Bézier function to ensure a smooth radial variation. 
The corresponding fit yields our final $V_{\rm rad}$ profile. 
As shown below, this procedure reliably recovers both radial velocities and warps in disc galaxies.

\end{itemize}

Since our approach departs from the default \texttt{$^{\rm 3D}$Barolo} fitting strategy, we do not rely on the default uncertainties provided by \texttt{$^{\rm 3D}$Barolo}, which are only approximate. Instead, we implement a dedicated error estimation method to obtain more representative uncertainties.
The error estimation is performed using a bootstrap approach \citep{2010MNRAS.404.1733S}. First, we compute the preliminary residuals by subtracting the best-fitting model datacube from the observed datacube. 
We then generate a new residual datacube by randomly permuting the spectral channels of the residual cube. This shuffled residual datacube is then added to the best-fitting model to produce a new mock datacube.
This process is repeated 50 times.
Each datacube is then refitted with our fiducial approach described above. 
Finally, we compute the mean and standard deviation of the fitted results. 
The final uncertainty is computed as the square root of the sum in quadrature of (i) the difference between the best-fit value and the bootstrap mean, and (ii) the standard deviation of the bootstrap distribution. 
As we show below, this approach yields realistic uncertainties.

\begin{figure*}[htbp] 
    \centering
    
    \includegraphics[width=0.98\textwidth]{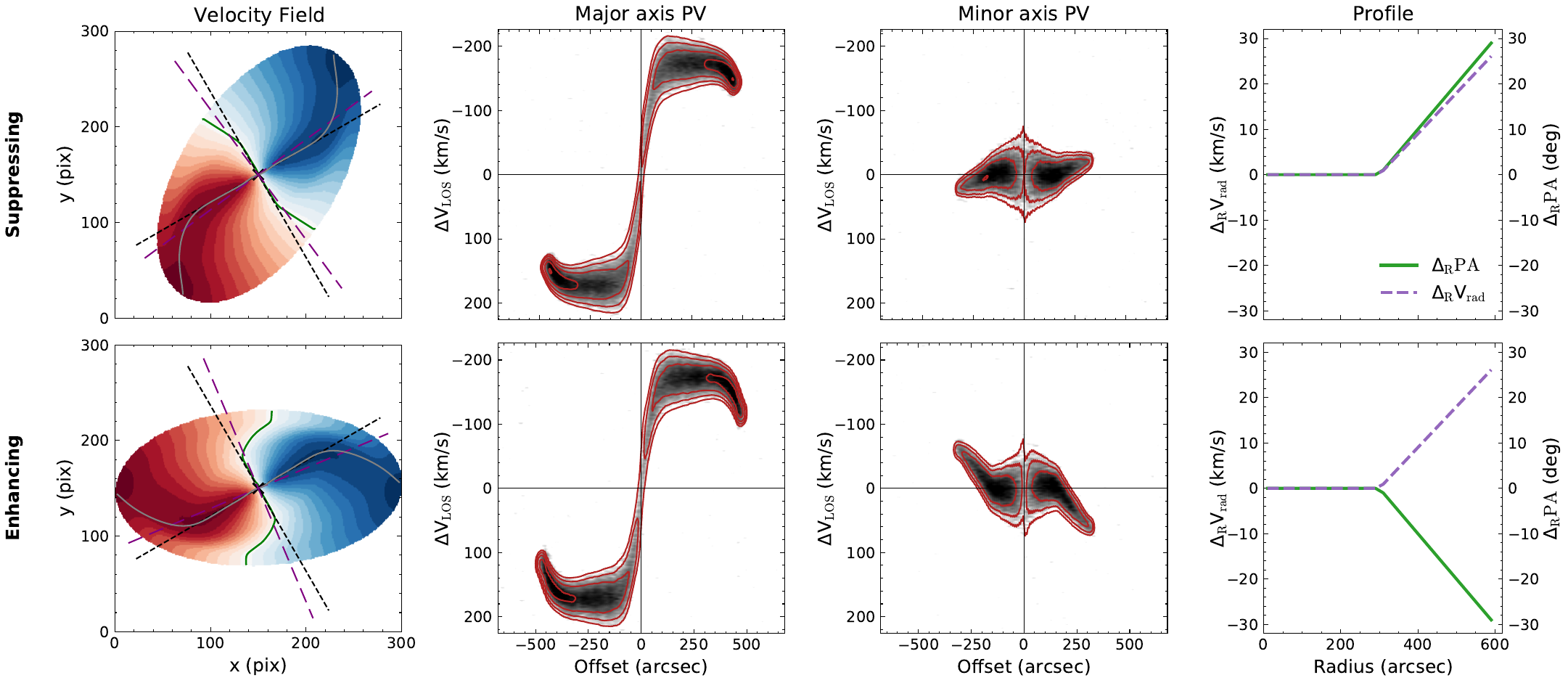}    
    \caption{
Toy models of a rotating disc with warps and/or radial motions, adopting a default inner position angle PA = $120^\circ$. The top and bottom rows show the warp--radial motion coupling in the suppressing and enhancing scenarios, respectively.
Relatively strong warp amplitudes are adopted here to emphasise the qualitative differences between the two kinematic configurations.
Left column: Line-of-sight velocity fields. 
The grey solid line shows the location of the highest projected velocities (kinematic major axis), while the green line marks the systemic-velocity contour. 
The black dashed lines show the major and minor axes of a purely rotating, unwarped disc, while the purple dashed lines indicate the average major and minor axes used for the PV extraction.
Middle columns: Position--velocity (PV) diagrams are taken from the standard \texttt{$^{\rm 3D}$Barolo} output and extracted along the average major and minor axes defined by the model PA, rather than the local ring-by-ring kinematic PA.
Red contours indicate iso-density levels.
Right column: Input radial profiles of the model parameters. 
The green solid line shows the radial variation of the position angle (PA), while the purple long-dashed line represents the radial velocity profile ($V_{\rm rad}$).
}

\label{fig:figure1}
\end{figure*}
\subsection{Construction of model galaxies}
\label{sec:model galaxies}
We first validate the method described in Sect.~\ref{sec:fitting_process} using two families of model datacubes, similar to \cite{2023ApJ...944..143W}:  
(a) warp-enhancing, where the warp structure amplifies the kinematic distortions produced by radial flows; and  
(b) warp-suppressing, where the warp structure reduces such distortions. 

To create the mock galaxies, we used the \texttt{GALMOD} task implemented in \texttt{$^{\rm 3D}$Barolo}, which was originally developed within the \texttt{GIPSY} software package \citep{1992ASPC...25..131V}. 
The grid size, spectral resolution, and beam size of the model galaxies resemble those of typical \hi observations of nearby galaxies, particularly those of the HALOGAS survey \citep{2011A&A...526A.118H}. 
To mimic nearby massive disc galaxies in the HALOGAS survey, we adopted a fiducial signal-to-noise ratio (S/N) of 10 and a spatial resolution of 30 independent resolution elements (beams) across the disc radius. 
The S/N is defined as the mean per-channel signal-to-noise ratio within the 3D mask constructed by \texttt{$^{\rm 3D}$Barolo}. Specifically, for each spectral channel, we compute the mean intensity within the mask, divide it by the per-channel rms noise (which is provided directly as input to \texttt{GALMOD}), and then average over channels with emission.
The choices of S/N values and spatial resolution ensure realistic observational conditions representative of well-resolved \hi discs in the local Universe.
We also investigated additional S/N values and spatial resolution configurations, as discussed below.

The disc kinematics and structure are defined by three primary profiles that extend up to $R=600''$. 
The rotation velocity follows a hyperbolic tangent curve, $V_{\rm rot}(r) = V_0 \tanh(r/R_{\rm rot})$, with $V_0 = 200$ km s$^{-1}$ and $R_{\rm rot} = 50''$. The velocity dispersion is described by an exponential profile, $\sigma(r) = \sigma_0 \exp(-r/R_{\sigma})$, where $\sigma_0 = 30$ km s$^{-1}$ and $R_{\sigma} = 500''$. 
The surface density is parametrised as $\Sigma(r) = \Sigma_0 (1 + r/r_*)^{\alpha} \exp(-r/r_*)$, adopting values of $\Sigma_0 = 2$, $r_* = 130''$, and $\alpha = 2.3$. The numerical values for these profiles were chosen to produce rotation curves, velocity dispersion profiles, and gas surface density profiles that are typically observed in nearby disc galaxies (e.g. \citealt{2021ApJ...923..220D}; \citealt{2025A&A...699A.311M}).

We further parametrise the position angle and radial velocity using a broken profile defined by a transition radius. Here we assume axisymmetric radial motions and model the warp as a variation of the PA in the outer parts of the disc. In Section \ref{Sec:inc_result} we also consider variations of the inclination angle, which makes the warp more realistic and generic (e.g. \cite{2006A&A...447...49B}). Inside this radius, ${\rm PA}$ and $V_{\rm rad}$ were fixed to $120^\circ$ and $0$, respectively. 
Beyond the transition radius (taken to be $300''$ or $450''$), both parameters varied linearly with radius, with slopes $\Delta_{\rm R}{\rm PA}$ and $\Delta_{\rm R}V_{\rm rad}$. We explored a range of radial gradients in the PA and $V_{\rm rad}$ profiles, with
$\Delta_{\rm R}{\rm PA}= [-0.1,-0.05,-0.025,0,0.025,0.05,0.1]$ in units of degrees per arcsec 
and
$\Delta_{\rm R}V_{\rm rad}= [-0.09,-0.07,-0.05,0,0.05,0.07,0.09]$ 
in ${\rm km}\,{\rm s}^{-1}$ per arcsec.
Over the full radial extent of the model, these gradients correspond to maximum variations of
$30^{\circ}$ in PA and $27~{\rm km\,s^{-1}}$ in $V_{\rm rad}$. Models were generated at fixed inclinations of $45^{\circ}$, $60^{\circ}$, and $75^{\circ}$, resulting in a baseline sample of 500 galaxies with an S/N value of 10 and 30 resolution elements.

The relative signs of $\Delta_{\rm R} \rm{PA}$ and $\Delta_{\rm R} V_{\rm{rad}}$ determine whether the warp structure enhances or suppresses the distortion in the velocity field driven by the radial motions. For outflows, $\Delta_{\rm R}V_{\rm rad} > 0$, negative values of $\Delta_{\rm R}{\rm PA}$ correspond to a warp-enhancing configuration, while positive values correspond to a warp-suppressing configuration. For inflows, $\Delta_{\rm R}V_{\rm rad} < 0$, this correspondence is reversed: positive $\Delta_{\rm R}{\rm PA}$ enhances the distortion, whereas negative $\Delta_{\rm R}{\rm PA}$ suppresses it.

To test the robustness of our method across different noise levels and spatial resolutions, we randomly selected 50 galaxies from the above sample and regenerated them with S/N $= 5$ and 20, and with spatial resolutions corresponding to 4, 7 and 15 independent beams across the galaxy radius, measured from the galaxy centre to the outer edge.
As mentioned, we generated galaxies at three fiducial inclinations to test how the inclination along the line-of-sight affects radial-velocity recovery. 
However, to consider possible warps in inclination, we also generated models with varying inclination as a function of radius.
For this purpose, we adopted the same transition radii mentioned above and introduced a linear inclination variation with slopes $\Delta_{\rm R} i = [-0.05, -0.025, 0.025, 0.05]$ in degrees per arcsec. 
The maximum inclination offset is $15^{\circ}$. 
To efficiently sample the full parameter space, we randomly selected 500 combinations to generate the corresponding datacubes. As in the fixed-inclination case, we randomly selected 50 galaxies from this sample and regenerated them with different S/N values and resolutions.

To illustrate the degeneracy described above, in Fig.~\ref{fig:figure1}, we present the velocity fields and position-velocity (PV) slices for two representative disc models with radial motions and PA variations in the suppressing and enhancing scenarios. The full set of toy models, including pure circular motions, radial motions with a constant PA, and PA warps without radial motions, is presented in Appendix~\ref{fig1appendix}.
The PV diagrams are taken from the standard \texttt{$^{\rm 3D}$Barolo} output and are extracted along the average major and minor axes of each model using the default WIDTH PV=0. This corresponds to a centre-line PV cut, approximately equivalent to a one-pixel-wide slice.
As the figure shows, both radial velocity and position-angle variations introduce distortions in the major and minor axes of the velocity fields, but can be highly degenerate due to the similar kinematic signatures.

\subsection{Construction of the mock observation of the TNG50 simulations}
\label{sec:buildTNGmocks}

While the mock galaxies constructed above provide controlled and realistic test cases in terms of kinematics, spatial resolution, and S/N values, idealised models do not fully capture the structural complexity and dynamics of real systems. 
In particular, they lack features such as asymmetric gas distributions, variations of gas turbulence across the disc, and effects of past interactions. 
To further test the robustness of our method under more realistic conditions, we therefore extracted disc galaxies from the TNG50 cosmological hydrodynamical simulation \citep{2019MNRAS.490.3234N,Pillepich+2019} and generated synthetic \hi datacubes from them.
The simulation box has a side length of 50 Mpc, with a maximum resolution of 70 pc. It is sampled by 2160$^3$ gas cells with a corresponding baryon mass of $8\times10^4\,M_\odot$ and dark matter mass of $5\times10^5\,M_\odot$.  
We focused on Milky Way and Andromeda analogues in TNG50 \citep{2024MNRAS.535.1721P}. 

We used the modular PYTHON package \texttt{Martini} \citep{2019MNRAS.482..821O} to generate mock \hi observations with realistic observing conditions. 
The \hi mass of gas cells was calculated with the \texttt{Hdecompose} function in \texttt{Martini}. 
We placed all galaxies at an arbitrary distance of 15 Mpc and generated datacubes with a spatial pixel size of $4''$. 
We adopted the observational setup of the HALOGAS galaxy NGC\,5055 \citep{2011A&A...526A.118H} to construct realistic mock datacubes. 
This corresponds to a beam size of $19.9'' \times 13.4''$ and a channel width of 4.1\,km\,s$^{-1}$; each cube contains 200 channels. 
To reproduce the noise properties typical of interferometric \hi observations, we injected realistic noise into the mock datacubes using a noise cube extracted from the HALOGAS data of NGC\,5055.
The final mock datacubes were obtained by linearly adding the noise cube to the beam-convolved model cube, allowing us to reach different S/N values. 
To test the effect of S/N on our method, we adjusted the noise level to achieve S/N values (averaged over all emission channels) of 5, 10, and 20. 
All mock observations were generated at a fixed inclination of \(i=60^\circ\) and a fiducial position angle of \(\phi=90^\circ\). 
We repeated the mock observations after rotating each galaxy within the disc plane by \(0^\circ\), \(90^\circ\), \(180^\circ\), and \(270^\circ\). 
This procedure is intended to reduce the impact of viewing direction relative to non-axisymmetric structures in the simulated gas discs.
Moreover, we smoothed the galaxies by factors of 2 and 4 relative to the beam size to obtain mocks at various spatial resolutions. The noise cube was smoothed to the same beam size prior to being added to the model cube. 

We selected highly symmetric disc galaxies based on the asymmetry metric defined by \cite{2026MNRAS.tmp..716H}. 
Specifically, we measured the 3D asymmetry by comparing each voxel in the datacube with its point-symmetric counterpart in both spatial and spectral dimensions, computing the normalised flux difference within a symmetric $\mathrm{^{3D}Barolo}$-based mask. 
We calculated the asymmetry parameter within the \hi radius $R_{\rm HI}$ defined by the region enclosed by the \hi column density corresponding to $1\,M_\odot/\rm{pc}^2$. For each galaxy, we measured this parameter at different S/N and spatial resolutions, and then adopted the average value as the representative asymmetry.
We retained galaxies with asymmetry parameter \(A\) below the selected percentile threshold of the asymmetry distribution, corresponding to the most symmetric 16\% of the sample. We focused on these relatively symmetric systems to minimise the impact of strong non-axisymmetric disturbances and to evaluate the performance of the method under well-behaved kinematic conditions.
We further selected galaxies through visual inspection, excluding the systems with bars (as often done in observational studies). 
After considering all these selection criteria, we obtained 12 TNG50 galaxies to test our method.

\section{Results}
\label{sec:results}

Here, we show the results of applying our new method to the model galaxies. In Sect. ~\ref{sec:model galaxies result}, we first focus on the models created with \texttt{GALMOD}, to then delve into the synthetic TNG50 galaxies in Sect. ~\ref{sec:resultsimulationgalaxies}.

\subsection{Application on model galaxies}
\label{sec:model galaxies result}
\subsubsection{Application on fixed inclination galaxies}
\label{sec:fixed_inc}

\begin{figure*}[htbp]   
    \centering

    \includegraphics[width=1\textwidth]{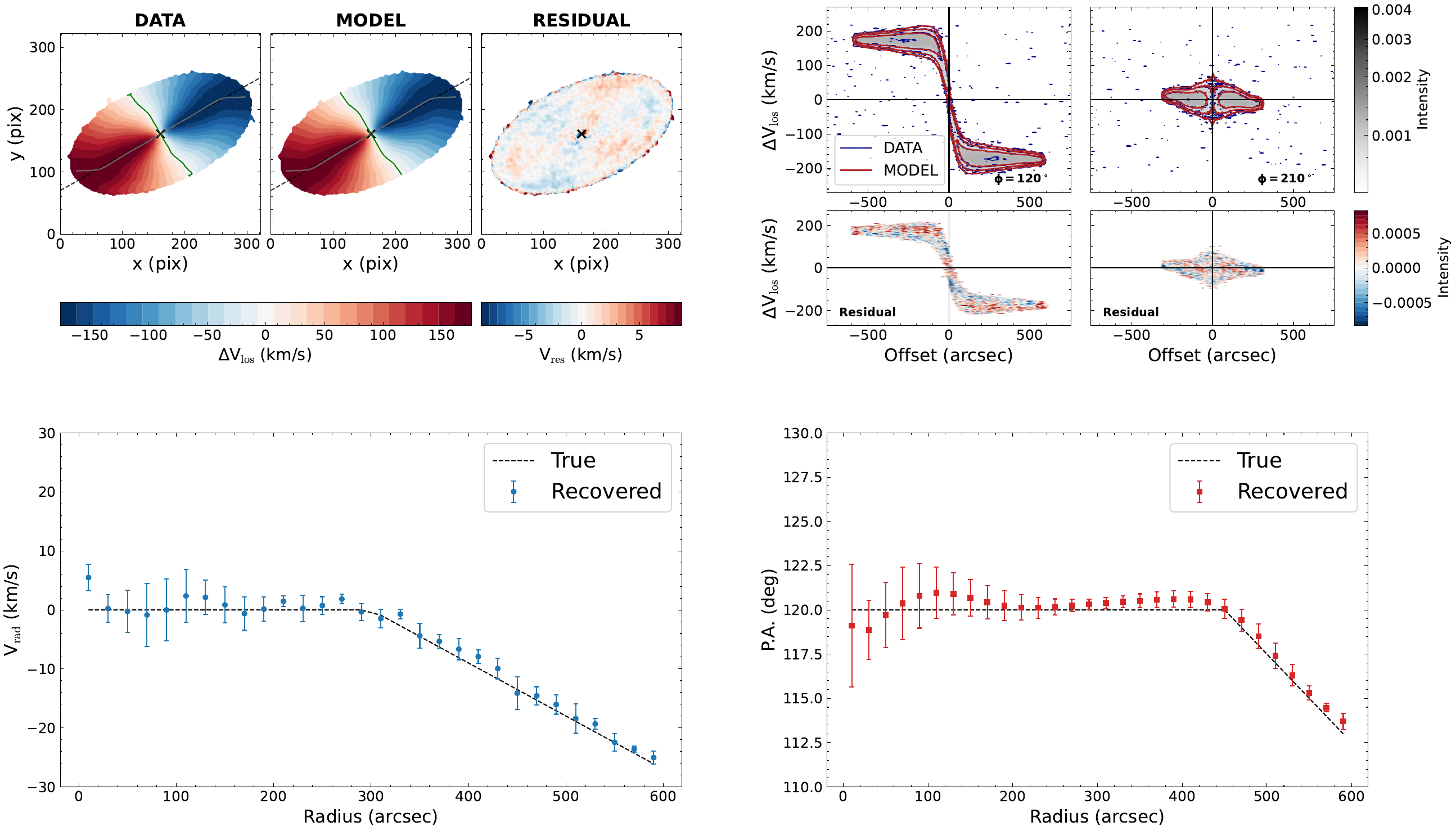}   
    \caption{
3D tilted-ring modelling of a model galaxy with a total $\Delta \mathrm{PA}=7.5\ \mathrm{deg}$ and a total $\Delta V_{\rm rad} = -27\ \mathrm{km\,s^{-1}}$. 
Top panels: left, line-of-sight velocity fields of the input model galaxy, the best-fitting tilted-ring model obtained with our approach described in Section~\ref{sec:fitting_process}, and the corresponding residual velocity field (data--model). 
Black crosses mark the adopted galaxy centre, green thick contours trace the systemic velocity ($\Delta V_{\mathrm{los}}=0$), and grey thin lines indicate the kinematic major axis. 
Right panels show position–velocity (PV) slices extracted along the average major and minor axes. 
The data (input model) and the recovered model are displayed as blue and red contours, respectively, with contour levels of $2^n\sigma_{\rm rms}$ ($n=0,1,2$). 
The lower panels display the corresponding PV residual intensity maps (data--model).
Bottom panels: comparison between the recovered kinematic profiles and the input model. 
The left panel shows the radial velocity profile, while the right panel presents the position angle as a function of radius.
}
\label{fig:singlegalaxy}
\end{figure*}

Figure~\ref{fig:singlegalaxy} illustrates our 3D tilted-ring modelling procedure described in Section~\ref{sec:fitting_process} and applied to a model warp-suppressing galaxy. 
The top panels compare the data and the best-fitting model for the line-of-sight velocity fields (left) and position--velocity (PV) diagrams extracted along the major and minor axes (right). 
The corresponding residual panels show the data--model differences in the line-of-sight velocity field and in the PV intensity distribution, providing a direct visual assessment of the fit quality.
The PV diagrams, in particular, demonstrate that the model reproduces the HI emission in the mock observation with high fidelity. 
The bottom panels show the $V_{\rm rad}$ and PA profiles recovered by our modelling procedure. 
In this case, the distortions in the velocity field induced by $V_{\rm rad}$ and PA variations partially counteract each other in projection. 
Despite this degeneracy, the method successfully recovers the underlying radial trends of both parameters.

\begin{figure*}[htbp]
    \centering
    \includegraphics[width=1\textwidth]{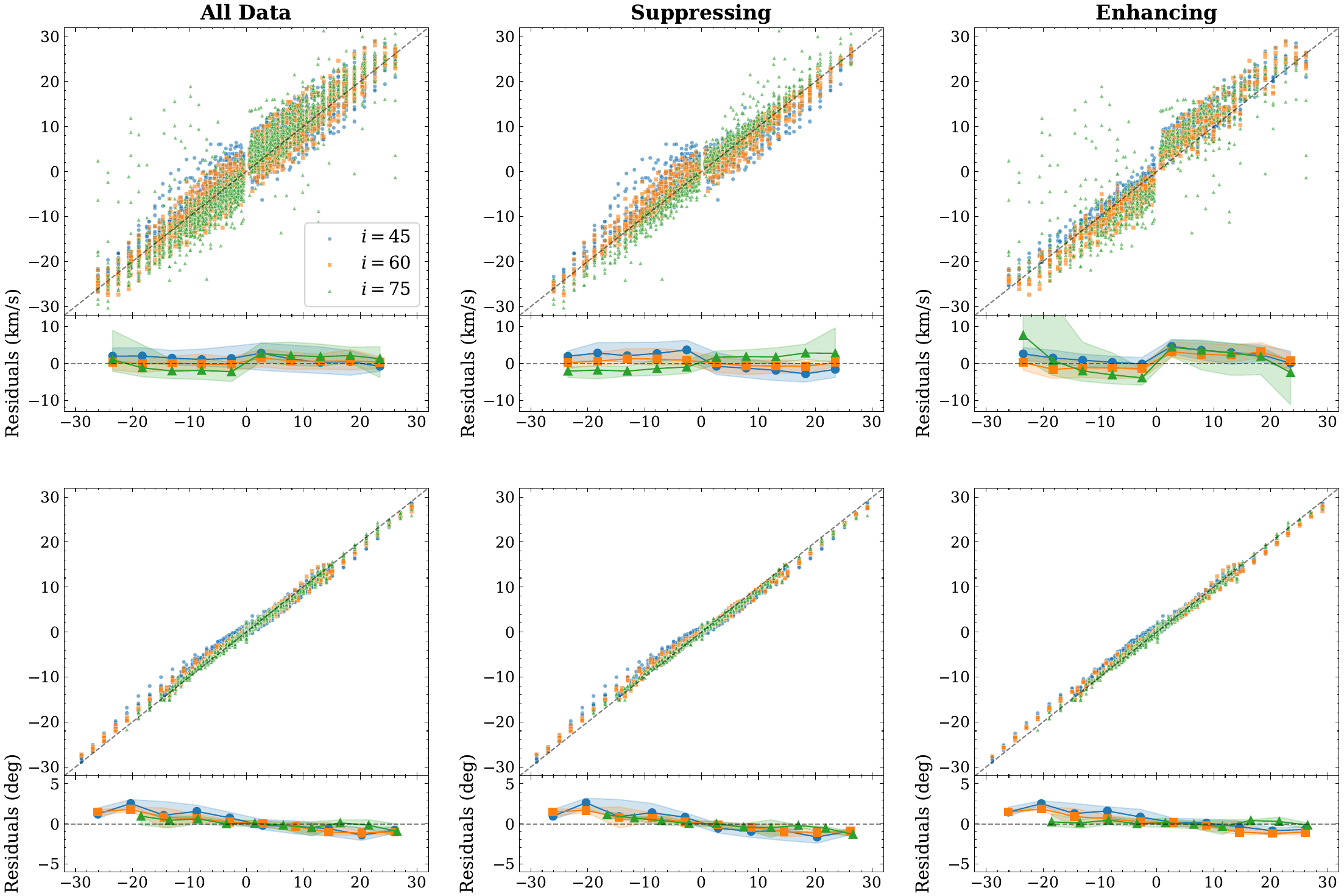}
    
    \caption{
        Comparison between the recovered kinematic parameters derived with our fitting procedure (Section \ref{sec:fitting_process}) and the input parameters. 
        Top Row: Radial Velocity ($V_{\mathrm{rad}}$). Bottom Row: Position Angle (PA). Columns (Left to Right): (1) All data points combined; (2) The warp-suppressing configuration; (3) The warp-enhancing configuration. 
        In each subplot, the upper panel shows the scatter plot of Recovered vs. True values, and the lower panel displays the binned residuals ($\text{Recovered} - \text{True}$).
        The data points are colour-coded by galaxy inclination: $i=45^\circ$ (blue circles), $i=60^\circ$ (orange squares), and $i=75^\circ$ (green triangles).
        The solid coloured lines in the residual panels indicate the median residual within bins, while the shaded regions represent the $16^{\rm th}-84^{\rm th}$ percentile range.
        The dashed black lines indicate the ideal 1-to-1 relation or zero residual.}

    \label{fig:recovery_comparison}
\end{figure*}

Figure~\ref{fig:recovery_comparison} presents the results of our method applied to all the model galaxies we created with fixed (i.e. radially constant) inclinations. 
The top row shows the recovery of the radial velocity, while the bottom row focuses on the recovery of the position angle. 
The three columns correspond to the full galaxy sample (left), the warp-suppressing configuration (middle), and the warp-enhancing configuration (right). 
In all panels, the residuals are defined as the difference between the recovered and true values. 
The shaded regions indicate the $16^{\rm th}-84^{\rm th}$ percentile range, and the galaxies are colour-coded by their inclination. 
All model galaxies shown in Fig.~\ref{fig:recovery_comparison} were generated at a fixed S/N $=10$ and a resolution of 30 resolution elements per side. 
Table~\ref{tab:rmse_values} provides a quantitative comparison of the root-mean-square error (RMSE) for the recovered kinematic parameters, computed for the full sample as well as for the warp-suppressing and warp-enhancing subsets.

\begin{table*}[htbp]
    \centering
    \caption{RMSE of the recovered radial velocities and position angles for different inclinations and kinematic configurations.}
    \label{tab:rmse_values}
    \setlength{\tabcolsep}{5pt}

    \begin{tabular}{l ccc c ccc}
        \hline\hline
        & \multicolumn{3}{c}{$V_{\rm rad}$ RMSE (km s$^{-1}$)}
        & & \multicolumn{3}{c}{PA RMSE ($^\circ$)} \\
        Inclination & All & Suppressing & Enhancing
        & & All & Suppressing & Enhancing \\
        \hline
        $45^\circ$ & 3.43 & 3.70 & 3.35 & & 1.20 & 1.29 & 1.16 \\
        $60^\circ$ & 2.63 & 2.20 & 3.20 & & 0.77 & 0.80 & 0.80 \\
        $75^\circ$ & 5.21 & 3.36 & 6.81 & & 0.60 & 0.62 & 0.63 \\
        \hline
    \end{tabular}
\end{table*}

We first focus on the recovery of the radial velocity. 
As shown in the top-left panel of Fig.~\ref{fig:recovery_comparison}, there is a great degree of agreement across the sample. 
Some inclination-dependent differences are visible, with the most accurate recovery obtained at the intermediate inclination of $i=60^\circ$, as also reflected in Table~\ref{tab:rmse_values}. For the \(i=75^\circ\) models, the recovered radial velocities show a small systematic bias away from zero, with the amplitude of \(V_{\rm rad}\) being slightly overestimated for both inflows and outflows. Nevertheless, the overall agreement remains good across the inclination range explored here.
We note that at low inclinations, the measurement of $V_{\rm rad}$ is fundamentally limited by projection effects. Since the line-of-sight component of the velocity scales with $\sin i$, the kinematic signature of radial flows becomes increasingly diminished in nearly face-on systems and is difficult to distinguish from observational noise. We also tested models with $i=20^\circ$, and, as expected, the recovery of $V_{\rm rad}$ becomes significantly less accurate under these conditions. An illustration of this test is shown in Appendix~\ref{app:lowinc}.
Conversely, at high inclinations, although the projection factor is maximised, the spatial resolution along the minor axis, where the $V_{\rm rad}$ signature is strongest ($\propto \sin\theta$), is degraded due to the compression of the projected minor-axis scale.
This geometry leads to strong line-of-sight integration and density overlap, preventing the tilted-ring model from azimuthally resolving the specific kinematic signatures of radial flows \citep{2015MNRAS.451.3021D}. 
As shown in the figure, highly inclined galaxies exhibit a larger scatter in the recovered $V_{\rm rad}$. 

We now turn to the recovery of the warps in position angle (we remind that we have constant inclinations here), shown in the bottom row of Fig.~\ref{fig:recovery_comparison}. 
The figure demonstrates that our method successfully recovers the PA profiles of the galaxies over the full range of input values. 
Upon closer inspection, however, a mild but systematic trend becomes apparent: negative warps tend to yield slightly overestimated PA values, whereas positive warps are modestly underestimated. 
Yet, the magnitude of this systematic effect remains small, with deviations typically below $\sim 2^\circ$.
For models with large negative warp amplitudes (e.g.\ total $\Delta \mathrm{PA} \approx -30^\circ$), the recovery accuracy degrades because the warped outer disc is projected, along the line of sight, onto regions of the inner disc. 
This spatial overlap of the rings causes emission from different radii to be blended in velocity space, thereby limiting the tilted-ring model's ability to recover the PA profiles.
In general, the PA recovery also depends on inclination. 
As the galaxy becomes more edge-on, the projected disc appears increasingly elongated, imposing stronger geometric constraints on the orientation of the major axis. 
As a result, PA recovery improves at high inclinations. 
This trend is opposite to that observed for the radial velocity, whose recovery relies primarily on minor-axis information that becomes increasingly unresolved at high inclinations.

The middle and right columns of Fig.~\ref{fig:recovery_comparison} highlight contrasting behaviours between the warp-suppressing and warp-enhancing configurations. In the warp-suppressing case (middle column), the recovered radial velocities are systematically lower (less pronounced) than the true inputs for $i=45^\circ$ and $60^\circ$. 
In contrast, the warp-enhancing configuration (right column) shows a systematic overestimation of $V_{\rm rad}$, as warps and radial flows constructively interfere to produce stronger distortions in the datacubes as shown by the velocity fields. 
Interestingly, for highly inclined systems ($i=75^\circ$) in the warp-suppressing configuration, the recovered $V_{\rm rad}$ exhibits a slight systematic positive bias rather than the damping observed at lower inclinations.

\begin{figure*}[htbp]
    \sidecaption
    \includegraphics[width=12cm]{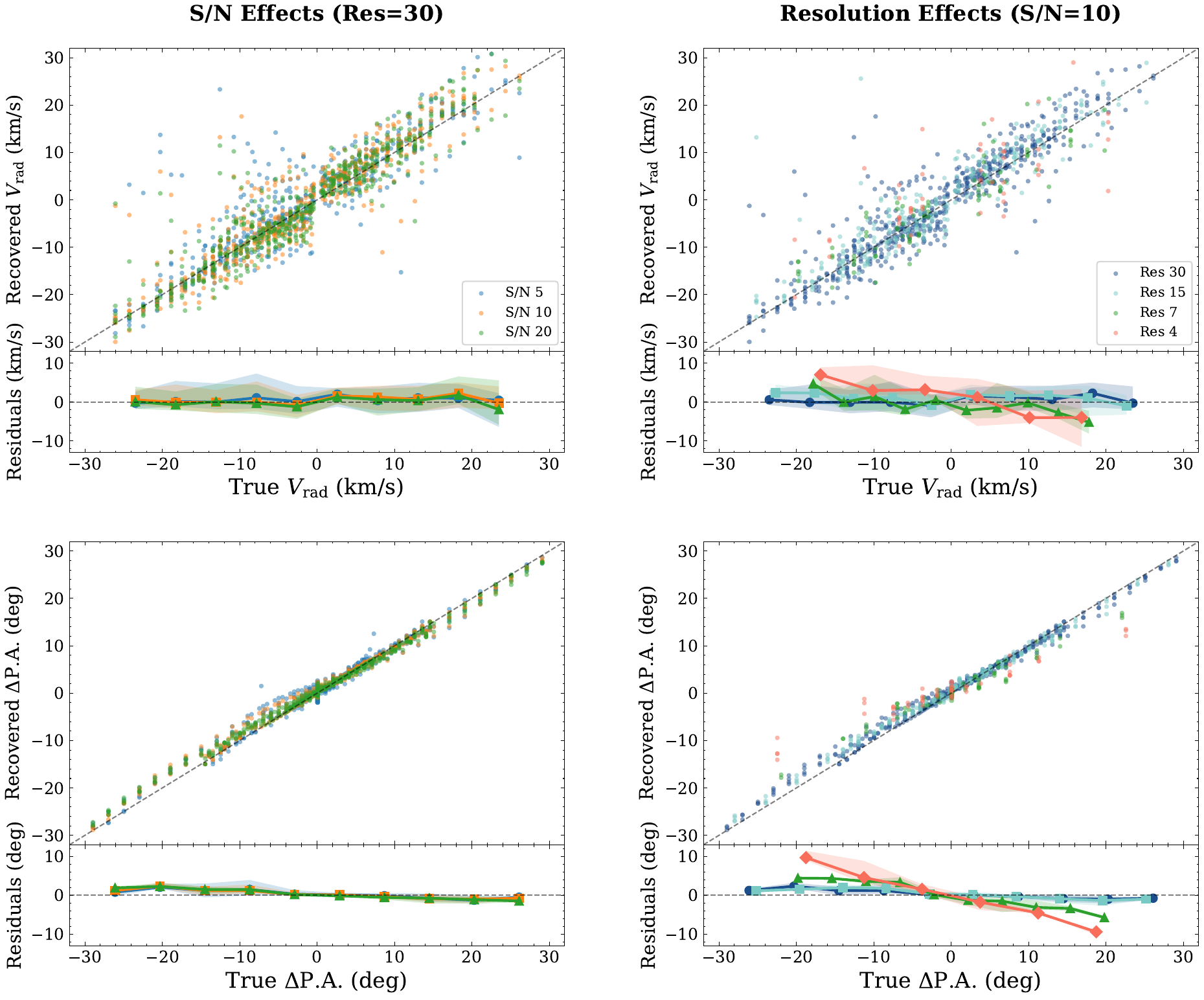}
    \caption{
    Recovery of kinematic parameters for mock galaxies across different S/N values and spatial resolutions; `Res' in the legend indicates the number of independent resolution elements per side of the disc.
    Colours indicate different S/N values and resolution configurations.
    The solid lines and shaded bands in the residual panels show the binned median residuals and the $16^{\rm th}$--$84^{\rm th}$ percentile range.
    }
    \label{fig:recovery_snr_resolution}
\end{figure*}
\subsubsection{Effect of the S/N and resolution}

To assess the performance limits of our method under varying observational conditions, we evaluated its sensitivity to S/N and spatial resolution. 
The model galaxies used in this test were randomly selected from the parent sample shown in Fig.~\ref{fig:recovery_comparison}, with models at all input inclination angles included. They were then re-generated with different S/N values and spatial resolutions, as described in Sect.~\ref{sec:model galaxies}.
In Fig.~\ref{fig:recovery_snr_resolution}, we show the dependence of the recovery accuracy on the S/N at fixed spatial resolution ($\mathrm{Res}=30$, i.e.\ 30 independent resolution elements), and on the spatial resolution at fixed S/N (S/N $=10$, average across all emission channels). 
Note that each point in Fig.~\ref{fig:recovery_snr_resolution} corresponds to a fitted ring, not to an individual galaxy. 

The left panel shows that, at high spatial resolution, the recovery of both \(V_{\rm rad}\) and PA is largely insensitive to S/N over the range explored, down to S/N = 5.
The median relation remains consistent with the one-to-one relation across the explored S/N range, indicating that the recovery is not systematically biased at low S/N provided that spatial resolution is sufficient.
The increase in scatter towards lower S/N is primarily driven by highly inclined galaxies, which are more sensitive to noise due to projection effects. 
This additional dispersion is present at all S/N values but becomes more pronounced as the S/N decreases.

Turning our attention now to the spatial resolution, we see that as it decreases from 30 to 4 resolution elements, the scatter increases. The recovery accuracy is comparable for $\mathrm{Res}=15$ and $\mathrm{Res}=30$, indicating that the method remains robust at moderate-to-high resolution. 
In contrast, at $\mathrm{Res}=4$, the scatter increases substantially, and the recovery starts to exhibit systematic deviations from the one-to-one relation. Because of the lower spatial resolution, the radial sampling is coarser, and each model galaxy therefore contributes fewer fitted rings to the recovered-versus-true comparison. In our mock galaxies, the radial variations in PA and radial velocity begin at different transition radii depending on the adopted model. At Res = 4, only a small number of fitted rings may therefore sample the radial range where the input parameters differ substantially from their inner values. As a result, the Res = 4 case contains fewer data points, particularly in the more extreme regions of the input parameter space. The regions containing only a small number of data points should therefore be interpreted with caution. The method not only shows greater dispersion but also begins to fail systematically at recovering the input parameters. This effect is particularly pronounced for the PA, where the bias becomes significant, and the deviations from the true values are no longer symmetric around zero. We inspected the $\mathrm{Res}=4$ cases and found that the low spatial resolution smooths out the kinematic distortions induced by radial flows, making $V_{\rm rad}$ more difficult to constrain. Moreover, the coarse resolution complicates the recovery of the velocity dispersion, which can increase the level of parameter compensation and further degrade the measurements of both $V_{\rm rad}$ and the position angle. Overall, these results demonstrate that our method works well across the tested S/N range and at moderate-to-high spatial resolution, but becomes less reliable when only four independent resolution elements are available.
This limitation should be carefully considered when applying the technique to nearby dwarf galaxies and high-$z$ systems, which can exhibit comparable spatial resolutions and may thus be subject to similar systematic effects.

\begin{figure*}[htbp]
    \centering
    \includegraphics[width=0.98\textwidth]{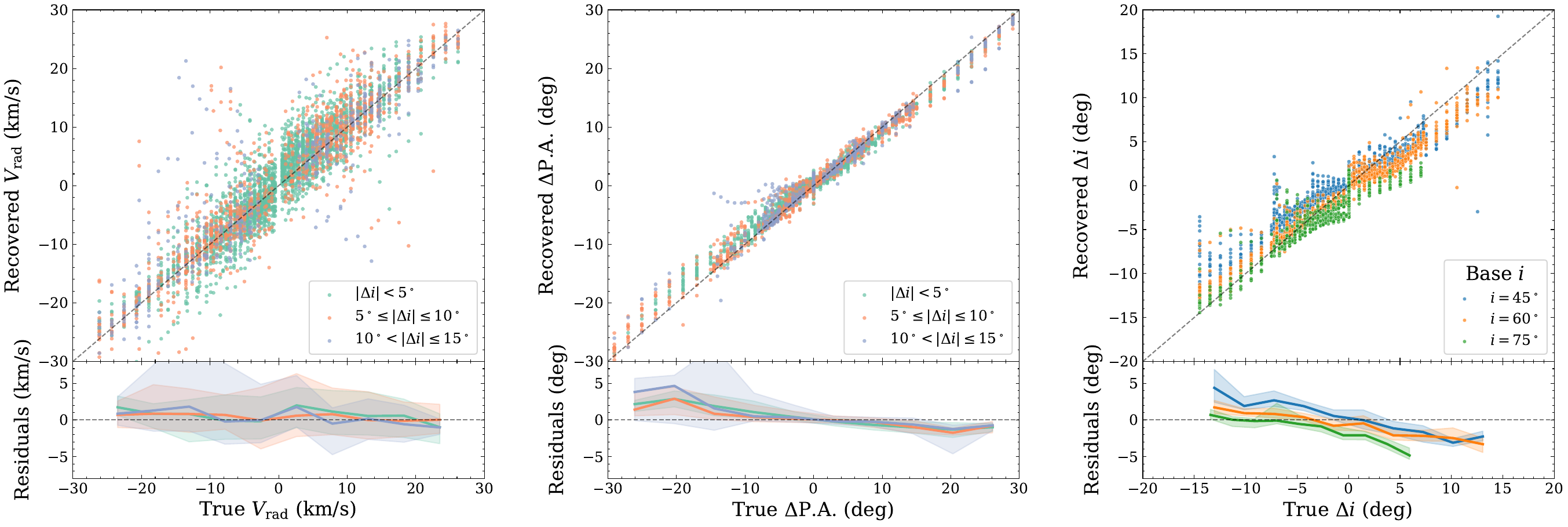}
    
    \caption{
Recovery of kinematic parameters for mock galaxies with warps in PA and inclination.
Columns from left to right: radial velocity ($V_{\rm rad}$), position angle (PA), and inclination offset ($\Delta i$).
In the $V_{\rm rad}$ and PA panels, points are colour-coded according to three bins of the absolute input inclination offset: $|\Delta i|<5^\circ$, $5^\circ\leq|\Delta i|\leq10^\circ$, and $10^\circ<|\Delta i|\leq15^\circ$.
In the inclination panel, points are colour-coded by the base input inclination.
The dashed black lines indicate the ideal one-to-one relations in the upper panels and zero residuals in the lower panels.
Solid lines in the residual panels show the binned median residuals for each subsample, while the shaded regions denote the corresponding $16^{\rm th}$--$84^{\rm th}$ percentile ranges.
We exclude nearly edge-on galaxies with input inclinations $i>80^\circ$.
      }

    \label{fig:recovery_varinc_vrad_pa_inc}
\end{figure*}

\subsubsection{Application on galaxies with varying inclination}
\label{Sec:inc_result}
In general, a galactic warp may not show up on the sky projection simply as a distortion of the PA of the outer rings, but also a change in inclinations \citep[e.g.][]{Battaglia+2006}. 
This can further increase the degeneracy between geometric effects and genuine radial motions \citep{2021ApJ...923..220D}. 
Here, we examine the recovery of geometrical angles and radial velocities for the set of mock galaxies in which the disc inclination varies with radius.

We applied the same four-step fitting strategy described in Sect.~\ref{sec:fitting_process} to this dataset, leaving the $V_{\rm rad}$ and PA as free parameters in the third step. 
We evaluated the recovery of $V_{\rm rad}$, PA and the adopted inclination profile as a function of inclination and warp strength. 
In Fig.~\ref{fig:recovery_varinc_vrad_pa_inc}, we present the recovery trends and the corresponding scatter with the same S/N $ =10$ and spatial resolution Res $ =30$ as in Sect.~\ref{sec:fixed_inc}.

Overall, the $V_{\rm rad}$ and PA profiles are well recovered for most of the mock galaxies, even in the presence of inclination warps. 
However, the scatter is larger than in the previously presented experiments, and there are some values that are clearly not recovered. The scatter increases with increasing \(|\Delta i|\), both for the recovered \(V_{\rm rad}\) and PA. Interestingly, even when the inclination offset is not well recovered, with $|\Delta i|>10^\circ$, the recovered $V_{\rm rad}$ remains comparatively stable. This indicates that the radial flow recovery is relatively robust against errors in the inferred inclination variation. The most accurate measurements are obtained for galaxies with inclinations around $60^\circ$, consistent with results for galaxies with fixed inclinations.
In particular, one galaxy shows a prominent deviation from the one-to-one relation in the middle (PA) panel. 
This outlier corresponds to a configuration in which emission from different radii overlaps along the line of sight, producing a multi-component structure in the position--velocity diagram that cannot be represented by a single tilted ring at that radius, which is a limitation of the tilted-ring modelling assumption (working ring by ring) adopted by \texttt{$^{\rm 3D}$Barolo}. 
Apart from this specific case, the fitted inclination profiles show a mild tendency to underestimate the amplitude of the input inclination change, particularly in the outermost rings where the surface density is the lowest. 
Nevertheless, the simultaneous presence of inclination and PA warps does not prevent the recovery of radial motions over the range explored here, but they increase the uncertainty in the recovered \(V_{\rm rad}\) and PA. 
This effect should be considered when interpreting measurements for individual galaxies with strong inclination variations.

\begin{figure*}[htbp]
    \centering
    
    \includegraphics[width=0.9\textwidth]{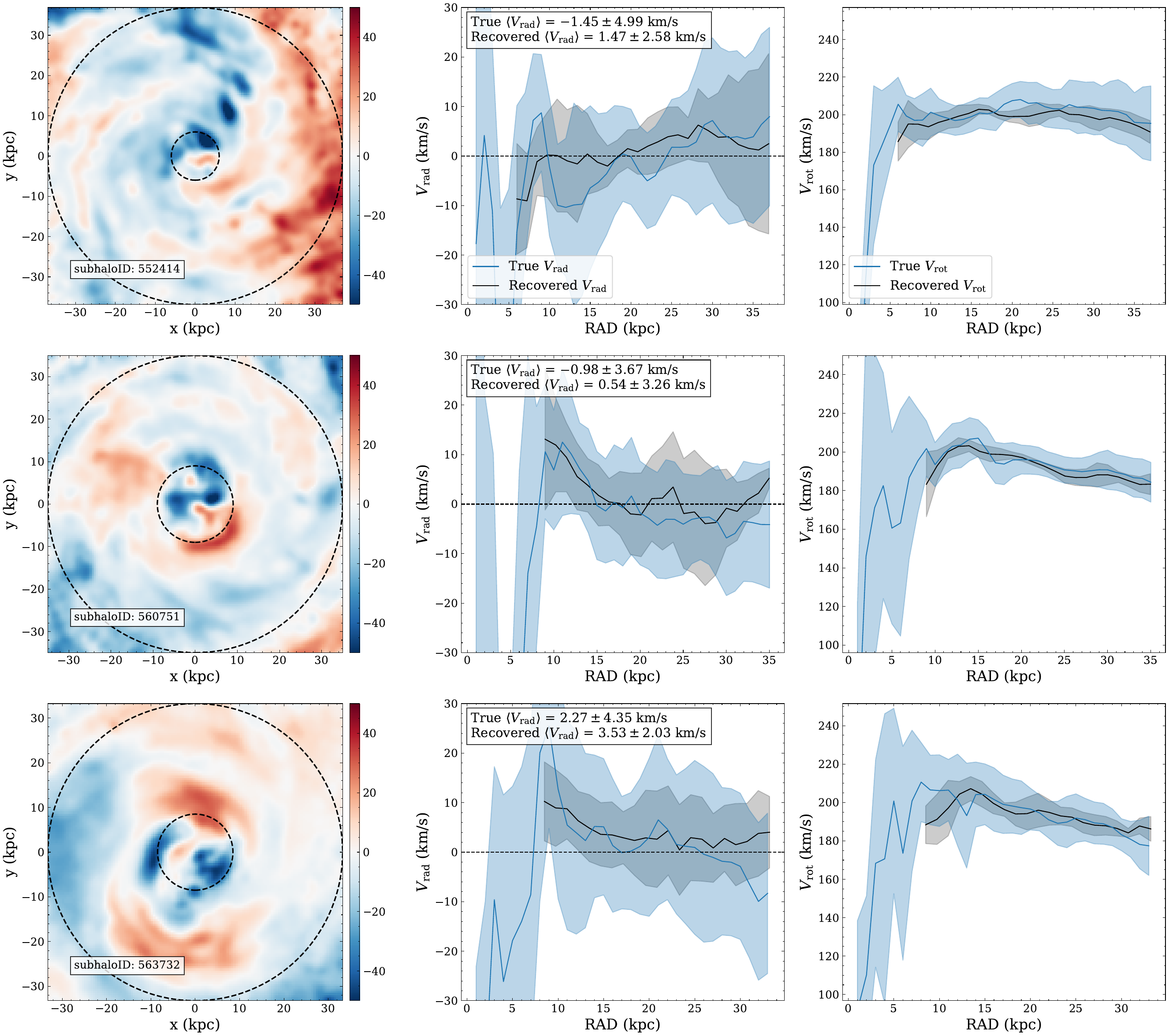}    
    \caption{Examples of three simulated galaxies illustrating the recovery of radial motions. The left panels show the intrinsic radial velocity fields of the gas in the face-on galaxy plane, with the colour bars indicating the radial velocity. The two dashed circles mark the inner and outer boundaries of the radial range over which the radial flow is measured, as defined by the radial extent of the \hi disc above a surface density of $1 M_\odot/{\rm pc}^{2}$. The middle panels show the recovered (black) and true (blue) radial velocity profiles as a function of radius. The horizontal dashed lines indicate $V_{\rm rad}=0$. The mass-weighted mean values of the true and recovered $V_{\rm rad}$, calculated over the radial range covered by the recovered profile, are also indicated in the middle panels. The right panels show the recovered (black) and true (blue) rotation velocity profiles as a function of radius. The blue shaded regions indicate the $1\sigma$ scatter of the gas particles, while the grey shaded regions indicate the $1\sigma$ scatter of the recovered results across the four viewing directions (see text).}
    \label{fig:threesimgalaxy}
\end{figure*}

\subsection{Application on the TNG50 simulated galaxies}
\label{sec:resultsimulationgalaxies}

As described in Sect.~\ref{sec:buildTNGmocks}, mock galaxies drawn from simulations provide a more realistic and stringent test bed for our method. We therefore applied our method to these mock \hi\ datacubes to further assess its performance under realistic conditions. For the simulated galaxies, we define the radial velocity as the in-plane velocity component along the galactocentric radial direction, with negative values indicating inward motion.

Fig.~\ref{fig:threesimgalaxy} presents three representative examples of the simulated disc galaxies described in Section~\ref{sec:buildTNGmocks}. The results for the remaining 9 galaxies in the sample are shown in Appendix~\ref{appendix:allsim}. To specifically test the recovery of genuinely non-zero radial motions, we additionally selected two TNG50 disc galaxies from the Milky Way and Andromeda analogue sample \citep{2024MNRAS.535.1721P} that exhibit clear global inflow signatures. These galaxies do not belong to the highly symmetric sample defined in Sect.~\ref{sec:buildTNGmocks} and are analysed separately in Appendix~\ref{app:strong_inflow}. For each galaxy, we show the intrinsic face-on distribution of radial velocity in the disc plane (left), the recovered radial velocity profile as a function of radius (middle), and the recovered rotation curve (right). In both cases, the recovered profiles are compared with the true values measured directly from the simulation.
The recovered profiles are computed as the median of the four projection directions (see details in Section \ref{sec:buildTNGmocks}), while the shaded regions indicate the scatter among these projections. The true $V_{\rm rad}$ and $V_{\rm rot}$ are derived from the gas particles directly. To quantify the overall radial motion, we also calculate the mass-weighted mean $V_{\rm rad}$ for both the true and recovered profiles over the radial range covered by the recovered profile. The corresponding mass-weighted standard deviations are used to characterise the scatter around these mean values. Since we only consider regions with \hi surface densities above $1\,M_\odot\,\mathrm{pc^{-2}}$, the central region is often not recovered as these galaxies often display central HI holes.

The face-on radial velocity maps reveal complex internal dynamics, characterised by spiral-like structures and velocity gradients indicative of localised perturbations within the galactic discs. 
The radial motions, ranging from $-40$ to $+40\,\mathrm{km\,s^{-1}}$, vary significantly with azimuth across the disc. 
While the true azimuthally averaged radial velocities fluctuate only slightly around the zero-velocity axis, the recovered profiles (black lines) successfully trace these trends beyond the central region. The mass-weighted mean recovered radial velocities are generally consistent with the corresponding true values within their standard deviations.
The recovered rotation velocities also show excellent agreement with the true values measured directly from the simulation. Over the radial range larger than \(0.5\,R_{\rm HI}\), the median absolute differences between the recovered and intrinsic profiles are \(3.76~{\rm km\,s^{-1}}\) for \(V_{\rm rad}\) and \(3.4~{\rm km\,s^{-1}}\) for \(V_{\rm rot}\).
These examples demonstrate that despite non-axisymmetric patterns in the radial velocity field, our method reliably recovers the underlying radial velocity profile within the uncertainties. The two galaxies with strong global inflow analysed in Appendix~\ref{app:strong_inflow} provide a complementary test in the presence of genuinely non-zero radial motions. In both cases, the recovered profiles reproduce the overall inflow signatures seen in the intrinsic velocity fields, and the mass-weighted mean recovered radial velocities are consistent with the corresponding true values within their standard deviations. In Appendix~\ref{appendix:allsim}, we also show the mock velocity fields and the corresponding best-fitting models for all simulated galaxies, with the recovered ring-by-ring kinematic PA overlaid.

\section{Discussion}
\label{sec:discussion}

\subsection{Comparison with previous 3D tilted-ring approaches}
\label{subsec:di}
\begin{figure*}[htbp]
    \sidecaption
    \includegraphics[width=12cm]{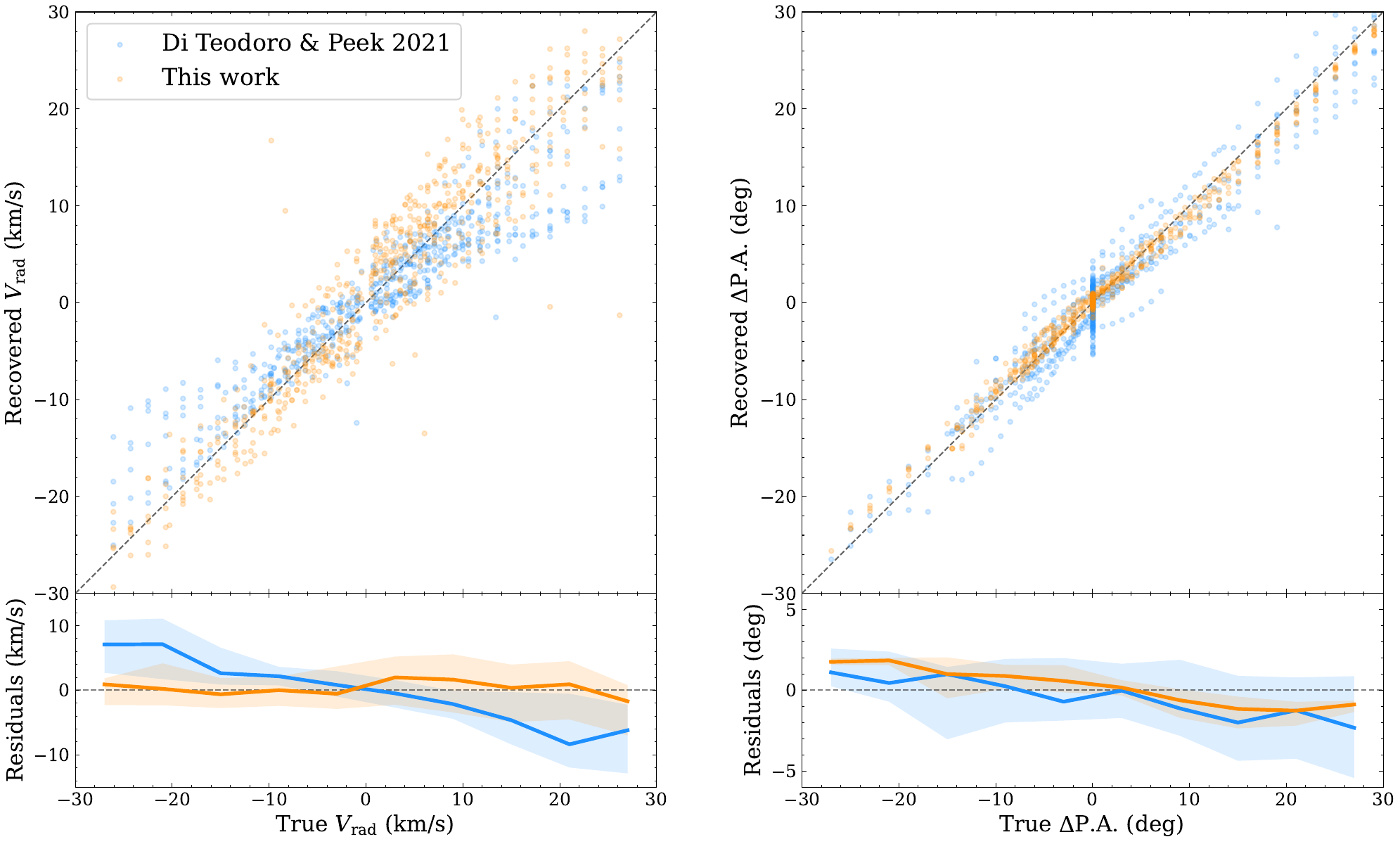}
    \caption{
    Comparison of $V_{\rm rad}$ (left) and PA (right) recovered with our method (orange) and with the approach of \citet{2021ApJ...923..220D} (blue).
    The dashed black lines mark the ideal one-to-one relation (upper panels) and zero residual (lower panels).
    Solid lines and shaded regions in the residual panels indicate the binned median residuals and the $16^{\rm th}$--$84^{\rm th}$ range.
    }
    \label{fig:comparisonwithenrico}
\end{figure*}

In \citet{2021ApJ...923..220D}, the authors compared their 3D tilted-ring approach with other commonly used techniques, including the classical 2D tilted-ring model of the velocity field \citep{1987PhDT.......199B}, Fourier decomposition methods \citep{2016MNRAS.457.2642S}, and the analysis of velocity profiles along the minor axis. 
Their study demonstrated that a 3D tilted-ring approach provides a more robust method for identifying radial motions in disc galaxies under realistic observational conditions. 
In their implementation, radial motions are set to zero ($V_{\rm rad}=0$) during the initial fit, while the geometrical parameters (inclination and position angle) and the other kinematical parameters are kept free. In a subsequent step, only the radial velocity is fitted, while all other parameters are fixed to the values determined in the previous steps.
However, since geometric distortions and radial motions are known to be partially degenerate in tilted-ring models, this sequential fitting strategy could bias how the signal is distributed between PA variations and $V_{\rm rad}$, as discussed in \citet{2023ApJ...944..143W} and \citet{2025ApJ...988..122S}. 
In contrast, to try to overcome such limitations, in our work, we fit the radial velocity and the warp structure simultaneously. 

In this Section, we perform a direct comparison between our method and that of \citet{2021ApJ...923..220D} in order to assess whether our simultaneous fit improves the recovery of the parameters. 
To perform this comparison, we randomly selected 60 model galaxies (Section~\ref{sec:model galaxies}) from our fixed-inclination sample, including $i = 45^\circ, 60^\circ ,75^\circ$ with S/N $=10$, Res $=30$, and applied both \citet{2021ApJ...923..220D}'s and our methods. 
The results of this comparison are shown in Fig.~\ref{fig:comparisonwithenrico}.

On average, both methods recover the true parameters. 
However, some systematic differences are noticeable in the radial velocity recovery.
With the method of \citet{2021ApJ...923..220D}, the recovered $V_{\rm rad}$ exhibits a systematic bias, with a general underestimation of the amplitude of the radial velocities. 
This trend is evident in the residuals as a monotonic deviation from zero with increasing $|V_{\rm rad}|$. As shown in Appendix~\ref{appendix:inc_di}, the strength and shape of this bias depend on the inclination, with a stronger effect for galaxies with inclinations $\lesssim60^\circ$.
These results indicate that fitting the warp geometry first can partially absorb the radial flow signal into the geometric parameters. 
In contrast, our method yields residuals that remain close to zero over the full range of tested radial velocities, without any strong biases.
The recovery of the position angle is robust for both methods, with tight clustering around the one-to-one relation. 
Nevertheless, the residuals from our simultaneous fitting approach show a slightly flatter trend and less scatter, suggesting weaker coupling between the geometric and kinematic parameters.

Overall, this comparison shows that while the 3D tilted-ring approach of \citet{2021ApJ...923..220D} is effective at identifying non-circular motions, simultaneously fitting the warp geometry and radial velocity yields a more accurate and less biased recovery of $V_{\rm rad}$. 
This improvement is particularly important at intermediate and low inclinations and, more generally, in systems where warps and radial motions coexist.
Given the ubiquity of \hi warps in local disc galaxies \citep[e.g.][]{1994A&A...290L...9R,2002A&A...394..769G}, this could be a realistic occurrence.

\subsection{Limitations of the method}
The method presented in this work is designed to recover the azimuthally averaged radial motions of gaseous discs using a tilted-ring model. As in most tilted-ring approaches \citep{1974ApJ...193..309R,1997MNRAS.292..349S,2010A&A...511A..89C,teodoro3DBaroloNew2015}, the gas kinematics are represented by a series of concentric rings, each characterised by a single set of geometrical and kinematic parameters. Consequently, the recovered radial velocity should be interpreted as the ring-averaged radial component rather than a full 2D description of local non-circular motions. The method is therefore specifically designed to recover axisymmetric radial flows. Local streaming motions associated with bars, spiral arms, lopsided gas distributions, or tidal interactions \citep{1992MNRAS.259..345A,1994A&A...290L...9R,1999ApJ...525..691S,2008AJ....136.2720T,2010MNRAS.404.1733S} are not explicitly modelled. These motions contribute to the measured radial velocity only if they produce a non-zero azimuthally averaged radial component within a ring. If radial transport is highly localised, for example along spiral arms or tidal features, the recovered radial velocity may not fully represent the instantaneous local gas flow. 

Moreover, the adopted tilted-ring model assumes a simplified vertical structure of the \hi\ disc. In particular, the vertical gas distribution is assumed to be fixed with radius, whereas real \hi\ discs are expected to flare towards large radii \citep{1995AJ....110..591O,1996AJ....112..481O,2008A&A...487..951K,mancerapinaImpactGasDisc2022}. Such flaring, together with possible vertical gas motions, may modify the projected kinematics in the outer disc. These effects are not included in the present model and may introduce additional uncertainties when analysing galaxies with pronounced vertical structure, in particular dwarf galaxies \citep{2019A&A...622A..64B}.

Another important assumption concerns the treatment of warps. In this work, warped discs are modelled through radial variations of PA and inclination. This provides a convenient way of reproducing the dominant observational signatures of many warped \hi\ discs and allows us to investigate the degeneracy between PA variations, inclination variations, and radial motions in a controlled manner. However, this treatment should not be regarded as a complete physical description of warped galaxies. Real warps correspond to 3D bending of the disc and may exhibit more complex geometries, including S-shaped or U-shaped distortions, different orientations of the line of nodes \citep{2002A&A...394..769G,2003A&A...399..457S,2006NewA...11..293A}. The present method is therefore expected to perform best with PA and inclination warps. More complex warp geometries may not be fully captured by the current implementation.

Overall, the tests presented in this work demonstrate that the method robustly recovers axisymmetric radial motions while simultaneously fitting PA variations, and remains reliable in the presence of inclination variations over the range of conditions explored here. Extending the method to treat more general 3D warp geometries and non-axisymmetric gas flows will be an important direction for future work.

\subsection{Radial flows in TNG50 Milky Way analogues}
\begin{figure}[htbp]
    \centering
   
    \includegraphics[width=\linewidth]{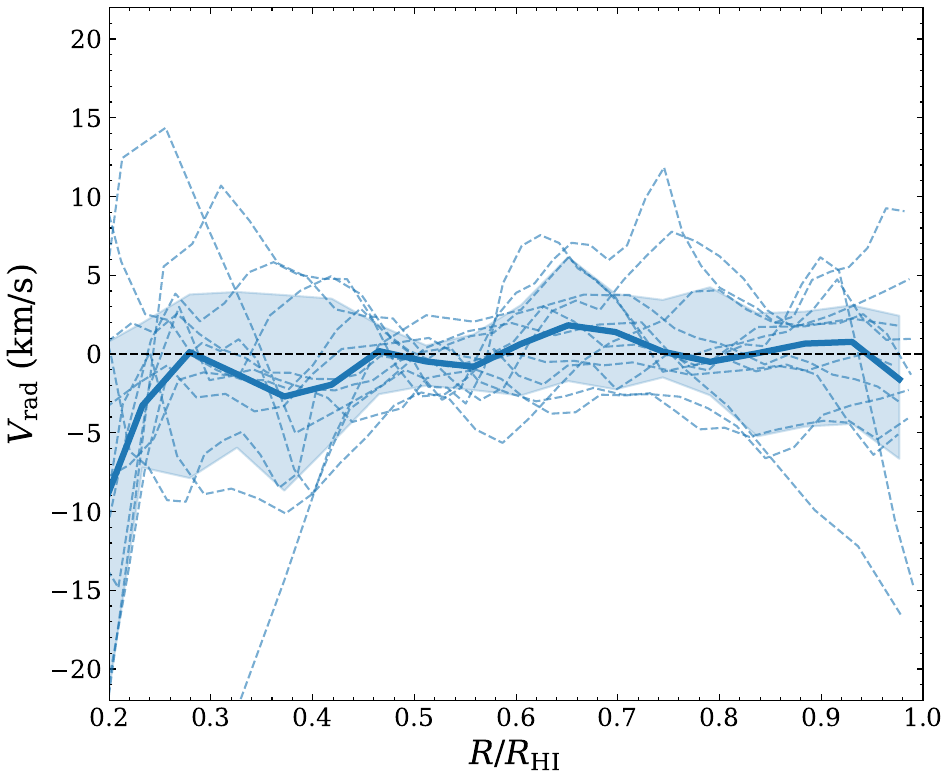}
    \caption{Radial velocity profiles of the simulated TNG50 galaxies. The horizontal axis shows the radius normalised by the \hi size, $R/R_{\rm HI}$, and the vertical axis shows the radial velocity $V_{\rm rad}$. The dashed blue lines represent the profiles of individual galaxies, while the solid blue line shows the median profile of the sample. The shaded region indicates the $16$–$84$ percentile range among the galaxies. The dashed black horizontal line marks $V_{\rm rad}=0$.}
    \label{fig:simvrad}
\end{figure}

Although our sample of simulated TNG50 galaxies consists of systems selected for their highly symmetric large-scale kinematics, the radial velocity field of the gas is not axisymmetric. Instead, the radial velocity varies with both azimuth and radius, as illustrated in Fig.~\ref{fig:threesimgalaxy}. 
To characterise the intrinsic radial motions in the simulations, we therefore computed the azimuthally averaged radial velocity directly from the gas particles in annuli. This provides a useful description of the radial velocity field and allows us to investigate the typical level of radial motions in simulated disc galaxies.

Figure~\ref{fig:simvrad} summarises the intrinsic radial velocity profiles of the simulated galaxies as a function of radius normalised by the \hi size, $R/R_{\rm HI}$. 
Overall, the median profile for all galaxies (thick solid curve) mildly fluctuates around $V_{\rm rad}=0$, showing no systematic trend with radius. 
Individual galaxies, however, can display larger local deviations, occasionally reaching radial velocities (both inflow and outflow) of $\sim10$ km/s at specific radii. The scatter around the individual dashed lines highlights the presence of localised kinematic perturbations, consistent with the azimuthal variations visible in the velocity fields shown in Fig.~\ref{fig:threesimgalaxy}. These fluctuations may be associated with internal dynamical structures such as spiral arms or bars. Ultimately, although our sample is limited, these results suggest that while local non-circular motions are present, the population as a whole does not exhibit coherent radial flows within the disc. 
This appears somewhat in contradiction with the results obtained by \citet{2022MNRAS.509.4149T} using four Milky Way-like galaxies extracted from the FIRE simulations, as those all show coherent inflow motions of, on average, $1-3\,{\rm km}\,{\rm s}^{-1}$ within their HI discs.
Despite both samples being relatively small, this difference in the two simulation suites could indicate different mechanisms of gas accretion and deserves further investigation (e.g. \citealt{2024MNRAS.532.3417W}).

\section{Conclusions}
\label{sec:conclusion}

We have presented a new methodology to measure axisymmetric radial motions in disc galaxies that employs a tilted-ring 3D fitting of emission-line datacubes. 
The main novelty of our approach compared to previous 3D methods 
is that radial motions are fitted simultaneously with the disc geometrical 
parameters, in particular the position angle, within the tilted-ring kinematic modelling implemented in \texttt{$^{\rm 3D}$Barolo}. 
In addition, we developed a statistical bootstrap procedure to obtain meaningful 
uncertainty estimates on the recovered kinematic parameters.

We tested the method on mock galaxies spanning a wide range of $V_{\rm rad}$, PA, inclinations, S/N values, and spatial resolutions typical of real \hi observations in the nearby Universe. Our method performs well across the tested S/N range and at moderate-to-high spatial resolution, successfully recovering both radial motions and position angle warps, but becomes less reliable when only four independent resolution elements per side are available.

Importantly, we show that fitting the warp geometry and $V_{\rm rad}$ simultaneously significantly mitigates the bias that can arise when the warp is determined first and then held fixed, as the signatures of significant radial inflows can otherwise be hidden within the warp measurements, particularly for inclinations $\leq 60^\circ$.

Using mock \hi datacubes constructed from the TNG50 hydrodynamic simulation, we further validated the approach under realistic observational conditions. 
We selected a sample of 12 Milky Way-like galaxies at \(z=0\) from the simulation and showed that we can recover radial profiles of $V_{\rm rot}$ and $V_{\rm rad}$ that agree well with their intrinsic profiles. 
Our analysis also indicates that the \hi radial velocities in these TNG50 disc galaxies exhibit strong azimuthal variations, with large regions in their discs displaying radial flows in both directions.
Once an azimuthal average is taken, there is no clear trend of the flows as the radial velocities average to zero across the discs.

In the future, we plan to apply the method presented here to an observational sample of nearby galaxies of different masses. This will allow us to (i) establish how common radial inflows are in nearby \hi discs, (ii) connect the inferred inflow rates to galaxy properties such as star formation rate, morphology, and environment, and (iii) constrain the relative importance of radial transport versus direct accretion onto the star-forming disc.\\

\begin{acknowledgements}

This work has received funding from the Program of China Scholarship Council (Grant No. 202204910131) and the European Research Council (ERC) under the Horizon Europe research and innovation programme (Acronym: FLOWS, Grant number: 101096087). PEMP is funded by the Dutch Research Council (NWO) through the Veni grant VI.Veni.222.364. We acknowledge the helpful discussions with Nathan Deg. We thank Barbara Šiljeg, Sten Sipma, and Jennifer Friske for insightful discussions and valuable feedback.
\end{acknowledgements}

\bibliographystyle{aa}
\bibliography{cite} 

\begin{appendix}
\section{Additional kinematic disc models}
\label{fig1appendix}
\begin{figure*}[htbp]
    \centering
    \includegraphics[width=0.98\textwidth]{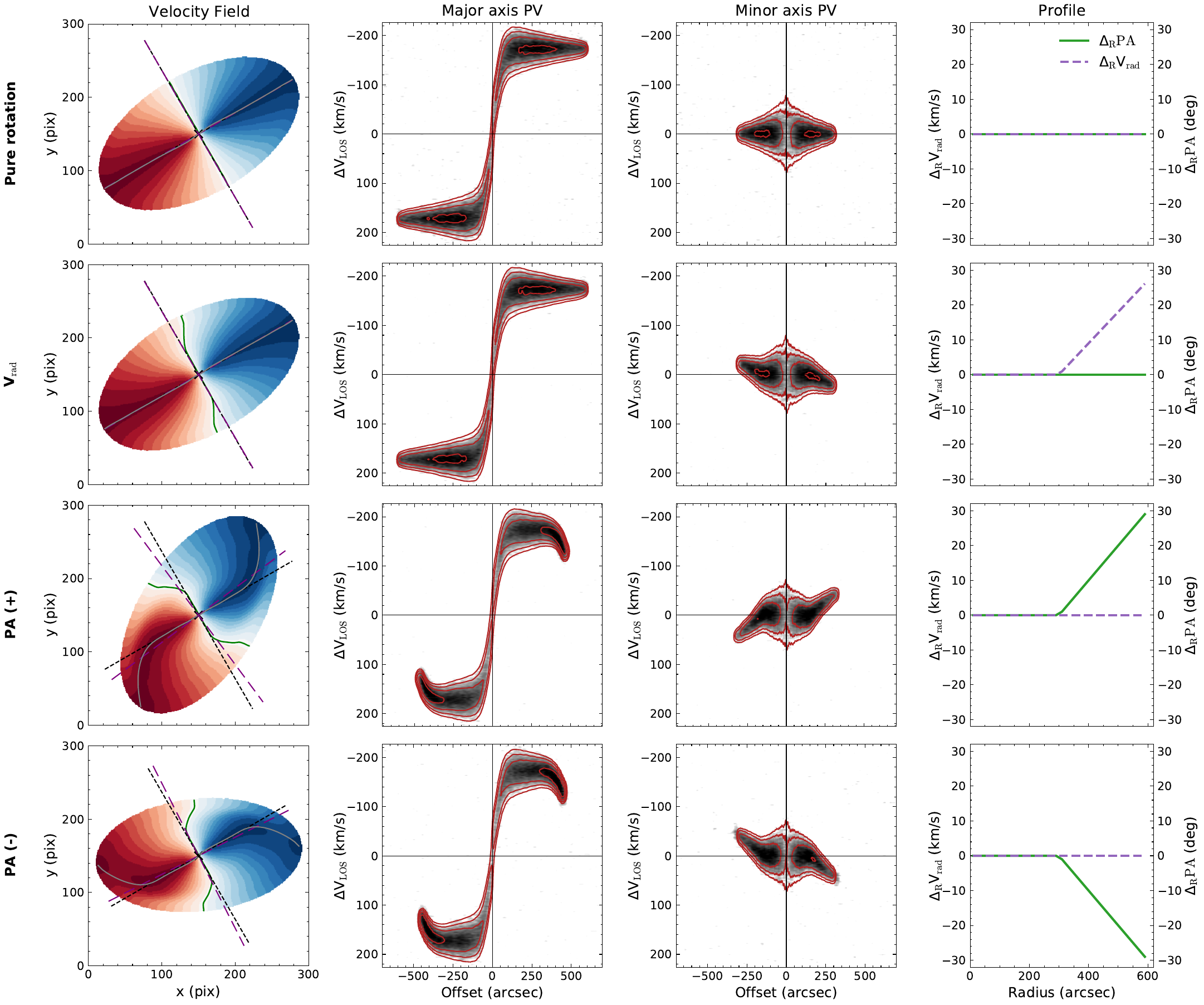}
 
    \caption{
Additional toy models of a rotating disc with warps and/or radial motions, adopting a default inner position angle PA = $120^\circ$. From top to bottom, the four rows correspond to (1) pure rotation; (2) radial flow with positive $V_{\rm rad}$ and a constant PA; (3) a warp with positive PA variation and no radial motions; and (4) a warp with negative PA variation and no radial motions. Relatively strong warp amplitudes are adopted here to emphasise the qualitative differences between the different kinematic configurations. The columns, lines, and contours are the same as in Fig.~\ref{fig:figure1}.} 
    \label{fig:appendixa1}
\end{figure*}

Figure~\ref{fig:appendixa1} presents additional kinematic disc models illustrating the individual effects of radial motions and PA variations. The models include a disc with pure circular motions, a disc with radial motions but a constant PA, and discs with positive or negative PA variations but no radial motions. Together with the suppressing and enhancing scenarios shown in Fig.~\ref{fig:figure1}, these models illustrate the different kinematic configurations considered in our analysis.

\section{The effect of low inclination on the $V_{\rm rad}$ recovery}
\label{app:lowinc}

\begin{figure*}[htbp]
    \sidecaption
    \includegraphics[width=12cm]{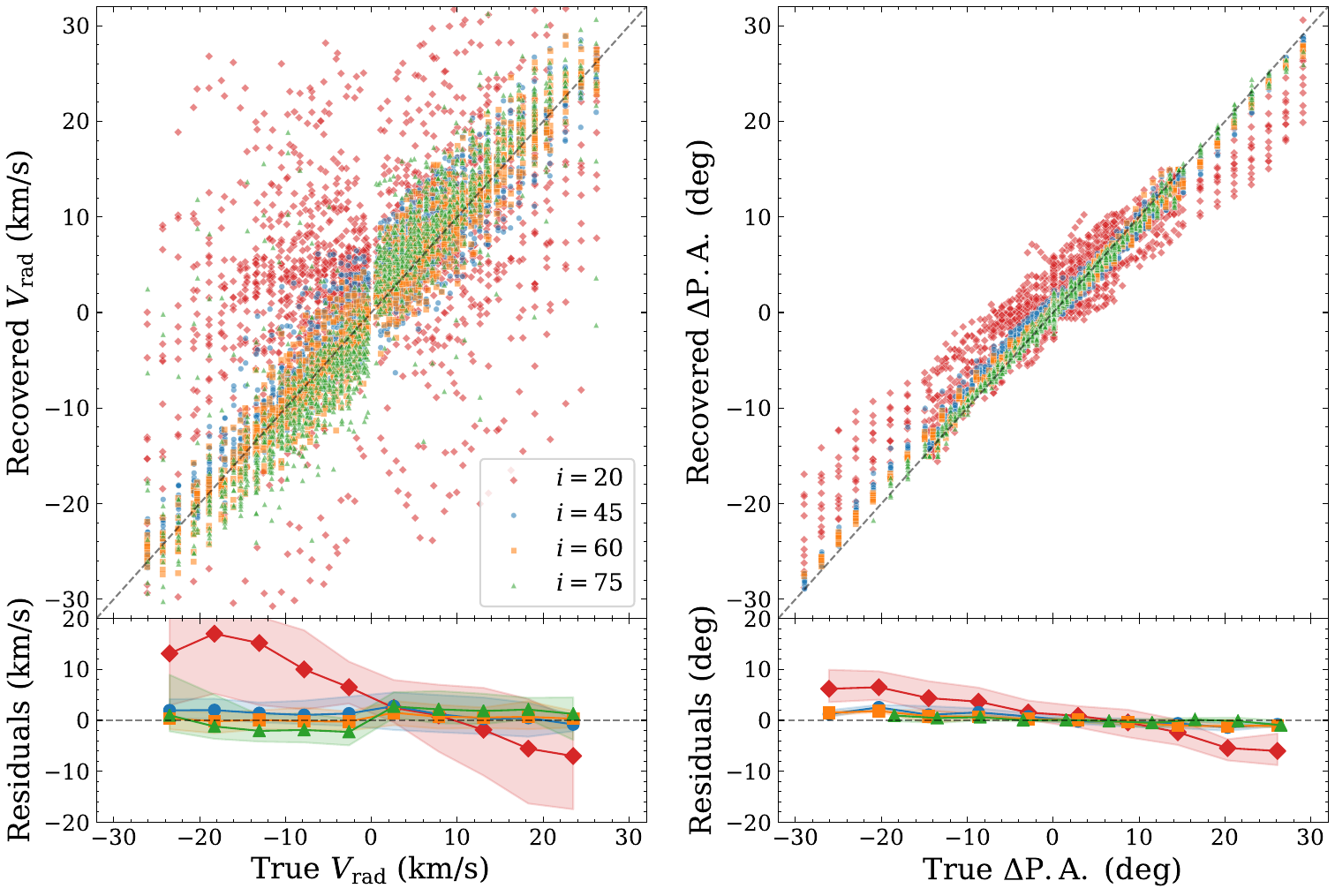}
    \caption{
    Comparison between the recovered kinematic parameters derived with our fitting procedure and the input parameters for mock galaxies with different inclinations, including the low-inclination case of $i=20^\circ$. In each subplot, the upper panel shows the recovered versus true values, while the lower panel displays the residuals (recovered $-$ true). The data points are colour-coded by galaxy inclination. The solid coloured lines in the residual panels indicate the median residual within bins, while the shaded regions represent the $16^{\rm th}$--$84^{\rm th}$ percentile range. The dashed black lines indicate the ideal one-to-one relation or zero residual.
    }
    \label{fig:lowinc}
\end{figure*}

Figure~\ref{fig:lowinc} shows the recovery of $V_{\rm rad}$ for mock galaxies with an inclination of $i=20^\circ$, compared with models at higher inclinations. As discussed in Sect.~3.1.1, the line-of-sight contribution from radial motions decreases as $\sin i$, making the recovery of $V_{\rm rad}$ increasingly difficult in nearly face-on systems. Compared to the higher-inclination models, the $i=20^\circ$ cases show substantially larger scatter and stronger deviations from the input values. This test therefore illustrates the practical limitation of measuring axisymmetric radial motions in galaxies viewed at low inclination.

\section{Application on all simulated galaxies}
\label{appendix:allsim}
In Fig.~\ref{fig:vradsim_all1}, we present the recovery results for all simulated galaxies in our sample, excluding those shown in Section~\ref{sec:resultsimulationgalaxies}. Each row corresponds to one galaxy. For each system, we show the face-on radial velocity map, along with the radial velocity profile and rotation curve as functions of radius. The recovered profiles are shown in black, while the true values measured directly from the gas particles in the simulations are shown in blue. In Fig.~\ref{fig:velocityfield}, we also show the mock velocity fields and the corresponding best-fitting models for all simulated galaxies. The black dashed line indicates the central kinematic PA, while the grey dashed line and the white circles trace the recovered ring-by-ring kinematic position angle. Several galaxies exhibit radial variations in the kinematic PA, particularly in their outer regions, while others show only mild PA variations. Overall, most of the recovered variations appear to follow regions of high line-of-sight velocity in the masked velocity field. This demonstrates that the method is capable of recovering kinematic PA profiles.

\begin{figure*}[htbp]
    \centering
    \includegraphics[width=0.9\textwidth]{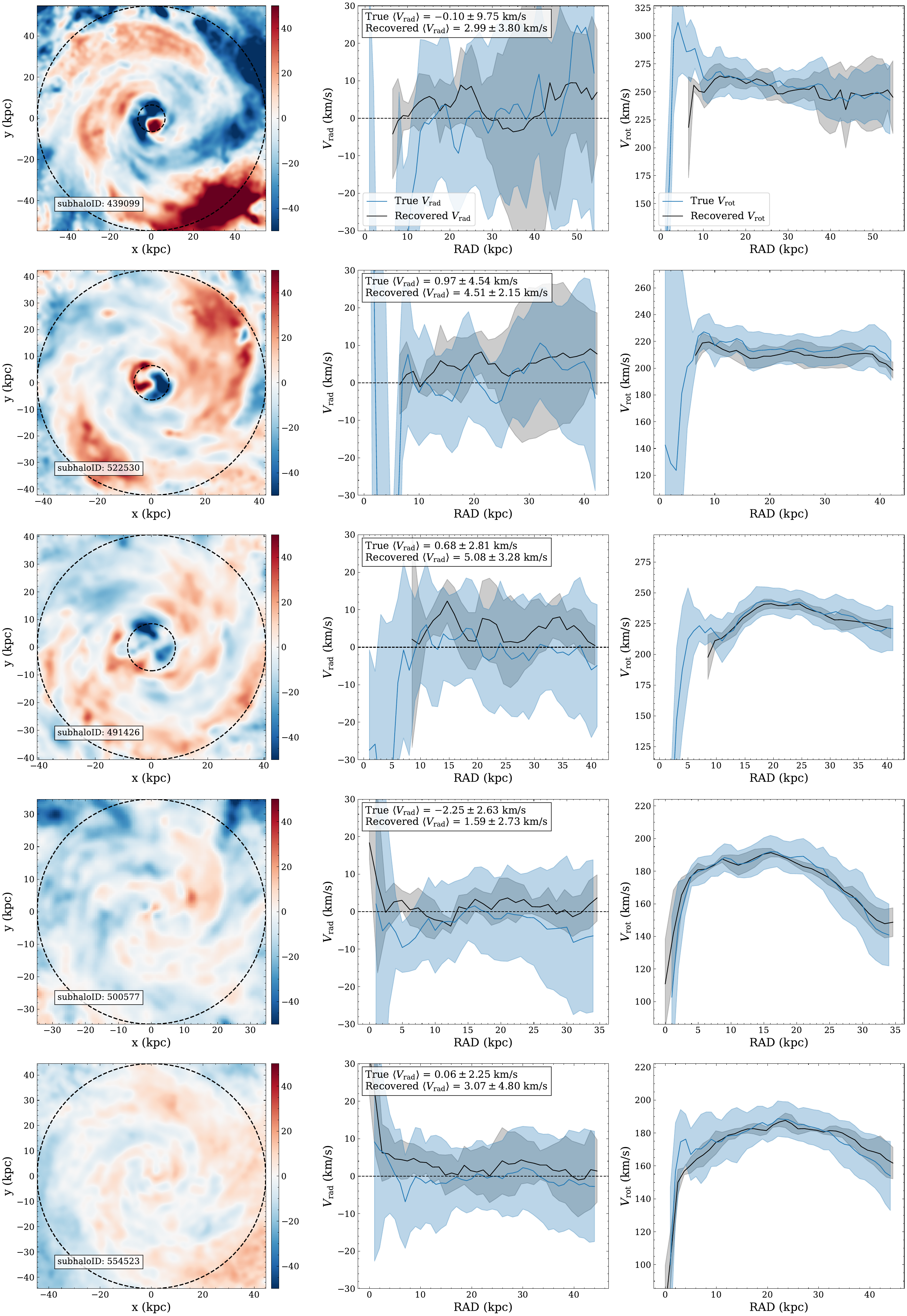}
    \caption{Kinematic recovery of simulated galaxies. Markers are the same as in Fig.~\ref{fig:threesimgalaxy}.}
    \label{fig:vradsim_all1}
\end{figure*}

\begin{figure*}[htbp]
    \centering
    \addtocounter{figure}{-1}
    \includegraphics[width=0.9\textwidth]{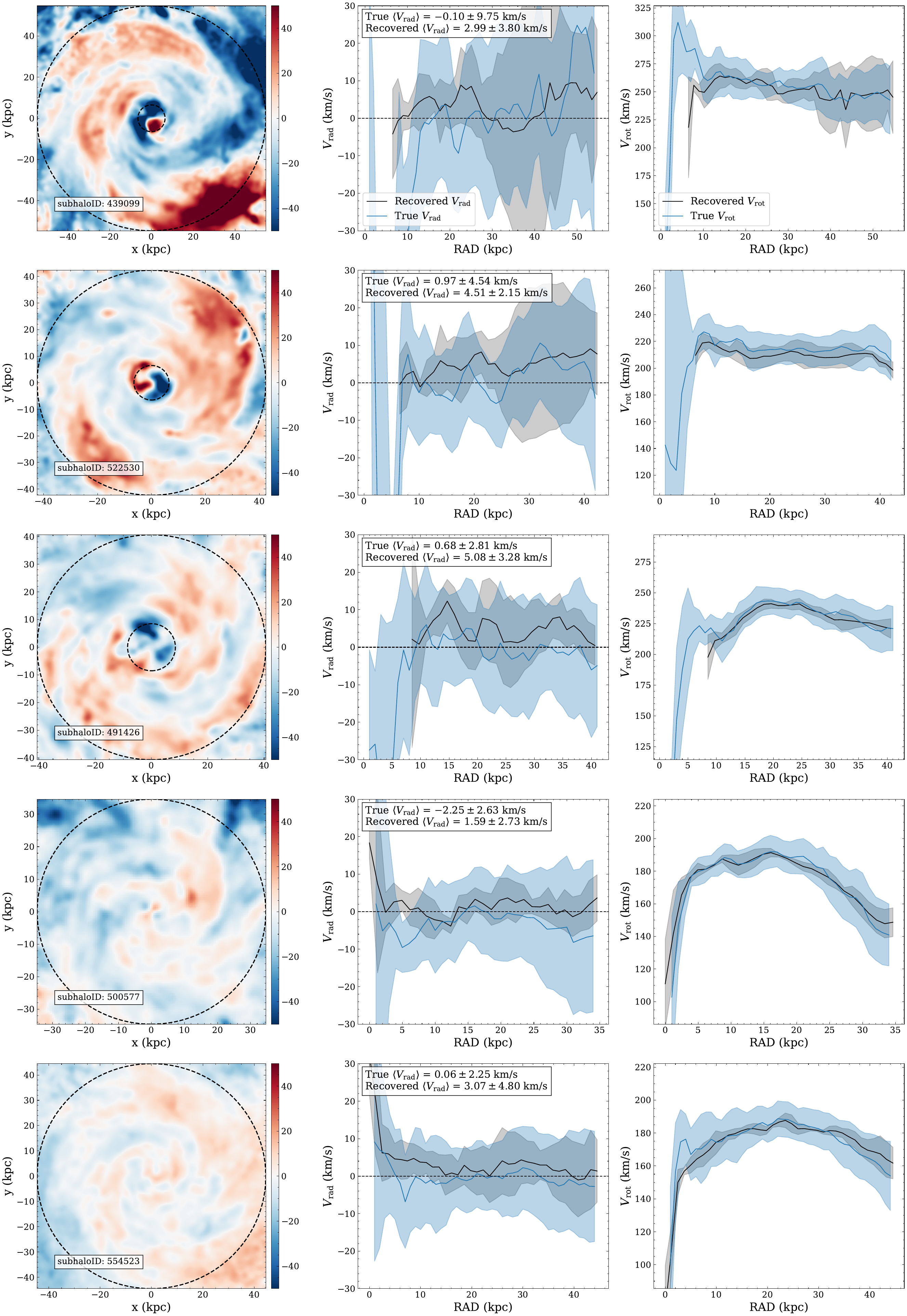}
    \caption{continued.}
    \label{fig:vradsim_all2}
\end{figure*}

\begin{figure*}[htbp]
    \centering
    
    \includegraphics[width=1\textwidth]{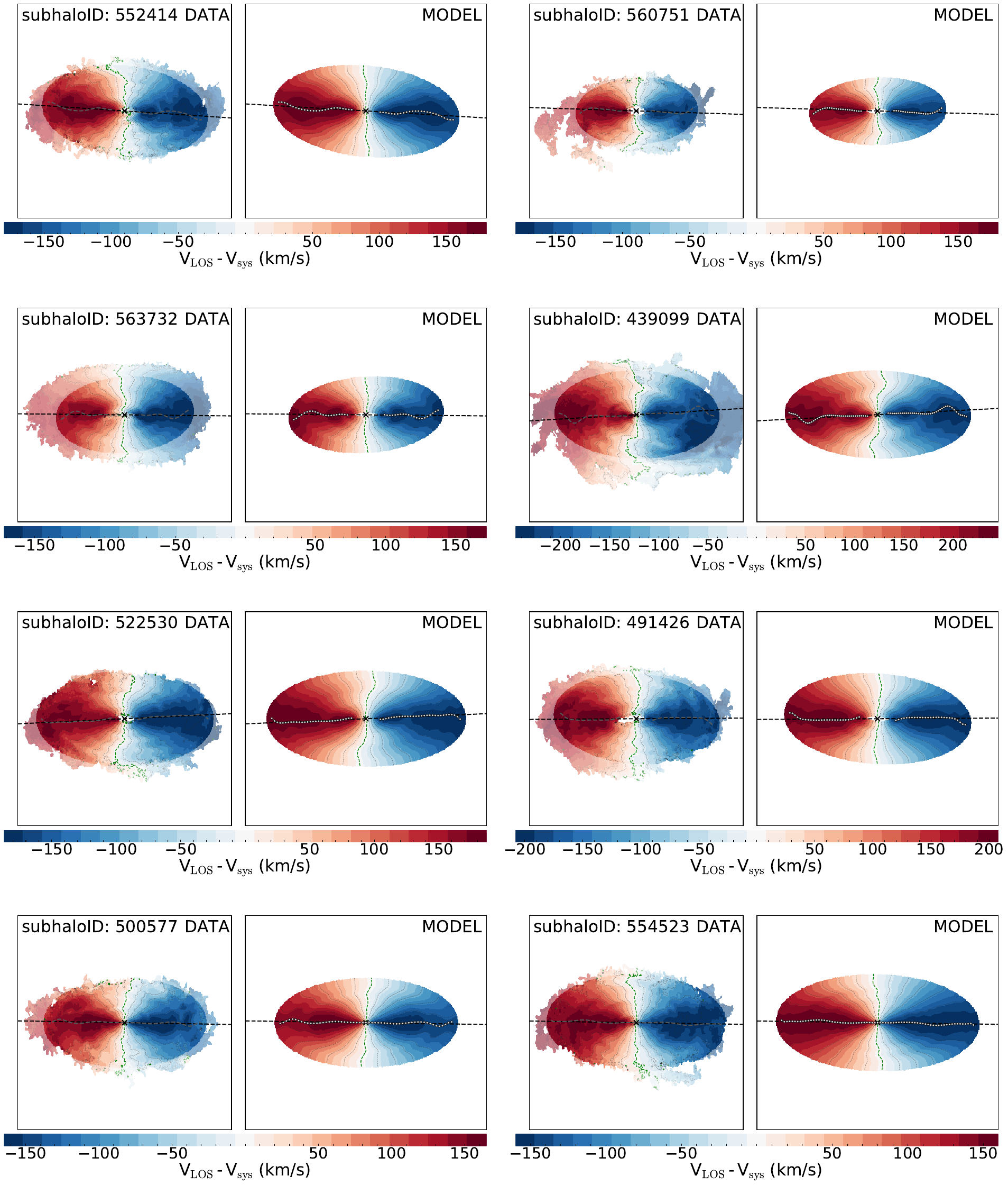}    
    \caption{Mock velocity fields (left) and the corresponding best-fitting models (right) for simulated galaxies. In the left panels, regions outside the H\,{\sc i} mask defined by the $1\,M_{\odot}$ threshold are shown with reduced opacity. The black dashed line marks the global kinematic position angle derived from the central disc, while the grey dashed lines and white circles show the fitted kinematic position angles.}
    \label{fig:velocityfield}
\end{figure*}
\begin{figure*}[htbp]
    \centering
    \addtocounter{figure}{-1}
    \includegraphics[width=\textwidth]{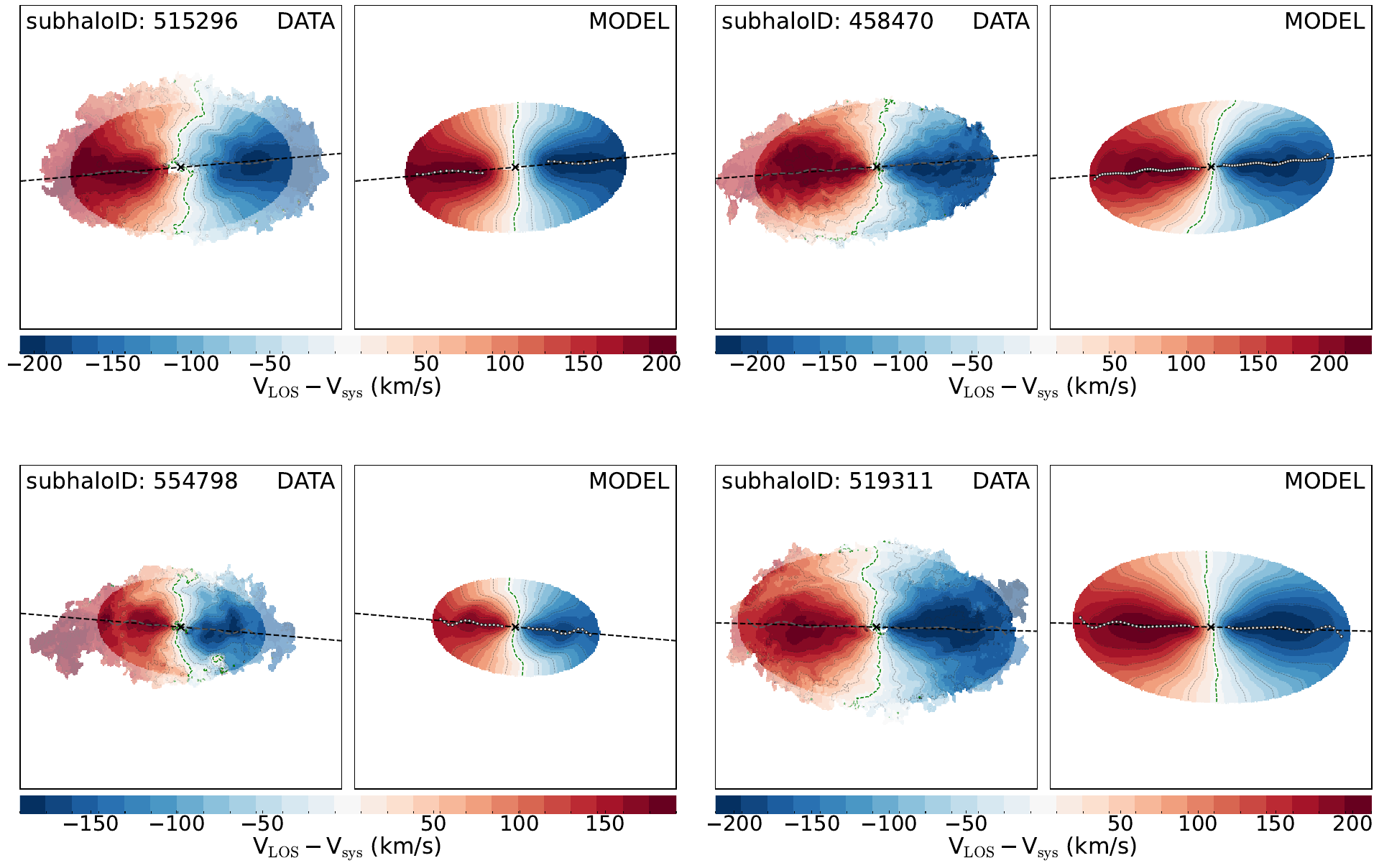}    
    \caption{continued.}
    \label{fig:velocityfield1}
\end{figure*}

\section{Recovery of strong radial inflows in simulated galaxies}
\label{app:strong_inflow}

The main simulation sample was selected to contain relatively symmetric and kinematically regular disc galaxies. As a complementary test, we selected two additional TNG50 galaxies that exhibit clear global inflow signatures. These systems do not satisfy the symmetry criterion adopted for the main sample and are therefore analysed separately.

\begin{figure*}[htbp]
    \sidecaption
    \includegraphics[width=12cm]{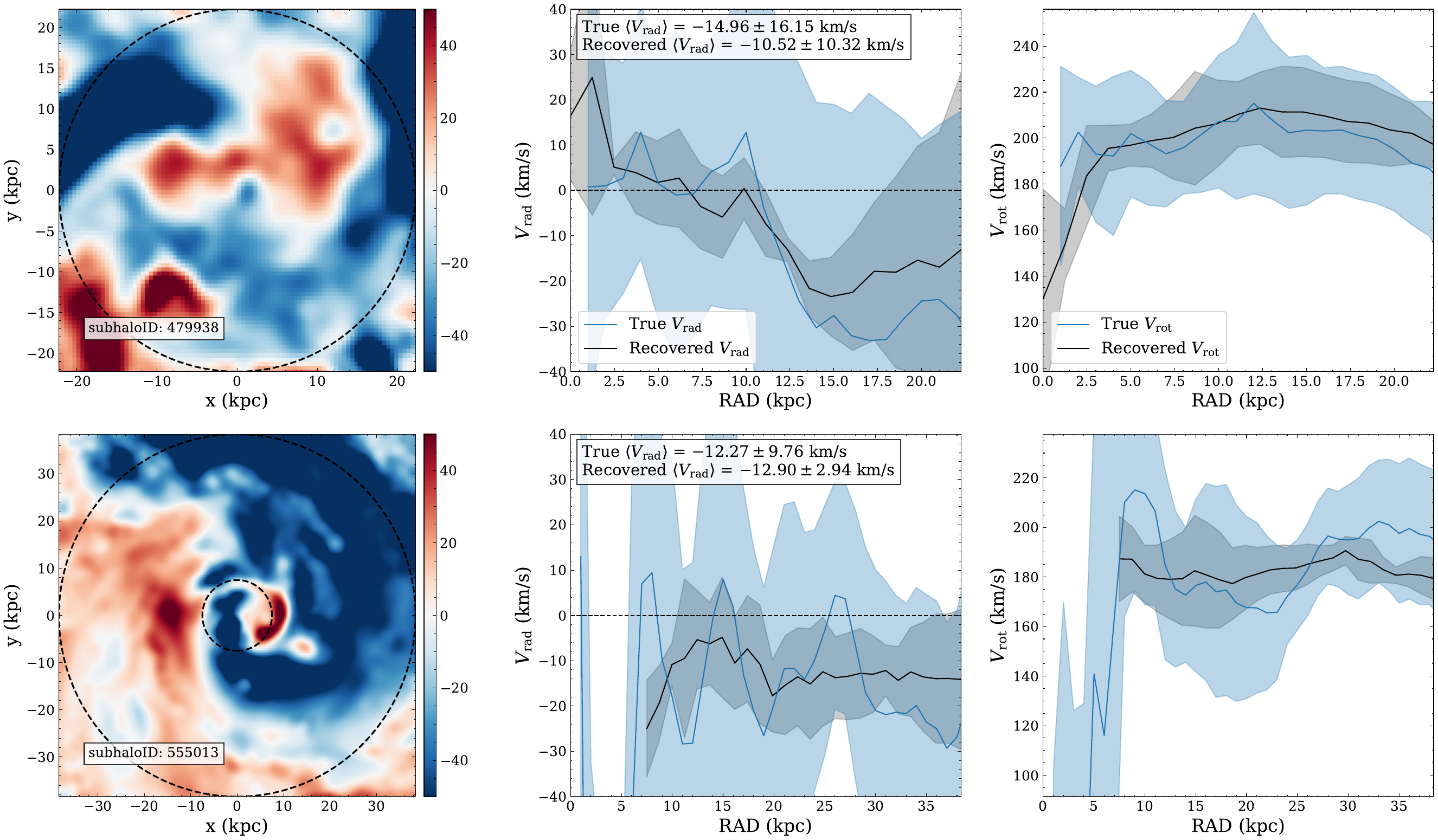}
    \caption{
    Kinematic recovery of two additional TNG50 galaxies exhibiting clear global inflow signatures. Markers are the same as in Fig.~\ref{fig:threesimgalaxy}.
    }
    \label{fig:twoinflowprofile}
\end{figure*}

\begin{figure*}[htbp]
    \sidecaption
    \includegraphics[width=12cm]{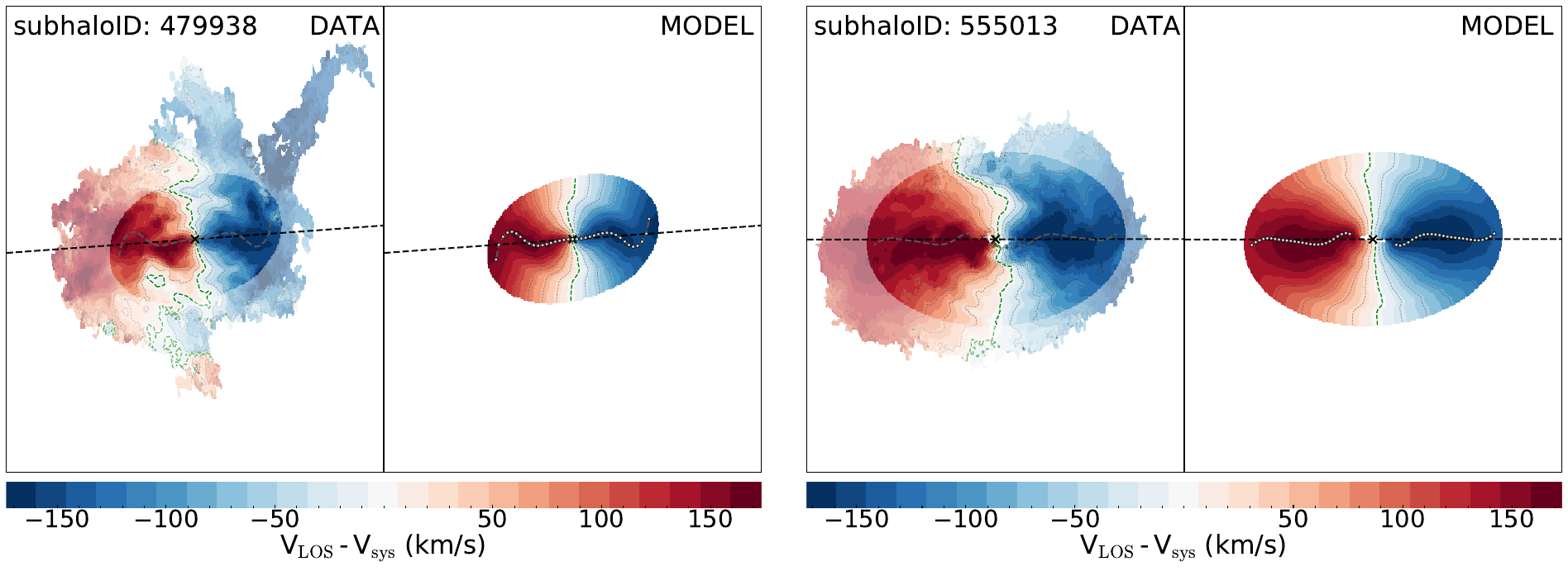}
    \caption{
    Mock velocity fields (left) and corresponding best-fitting models (right) for two additional TNG50 galaxies exhibiting clear global inflow signatures. Markers are the same as in Fig.~\ref{fig:velocityfield}.
    }
    \label{fig:twoinflowvelocityfiled}
\end{figure*}

Figure~\ref{fig:twoinflowprofile} compares the intrinsic and recovered radial velocity profiles for these two galaxies, while Figure~\ref{fig:twoinflowvelocityfiled} shows their corresponding velocity fields and best-fitting models. In both cases, the recovered profiles reproduce the overall negative sign and radial trend of the intrinsic $V_{\rm rad}$, demonstrating that the method is able to identify a genuine global inflow signal when present. Localised variations in the intrinsic profiles are not always reproduced in detail, as expected given that the tilted-ring model primarily constrains the axisymmetric component of the radial velocity field.


\section{The effect of the weighting scheme on the $V_{\rm rad}$ recovery}
\label{sec:weighting scheme}

\begin{figure*}[htbp]
    \sidecaption
    \includegraphics[width=12cm]{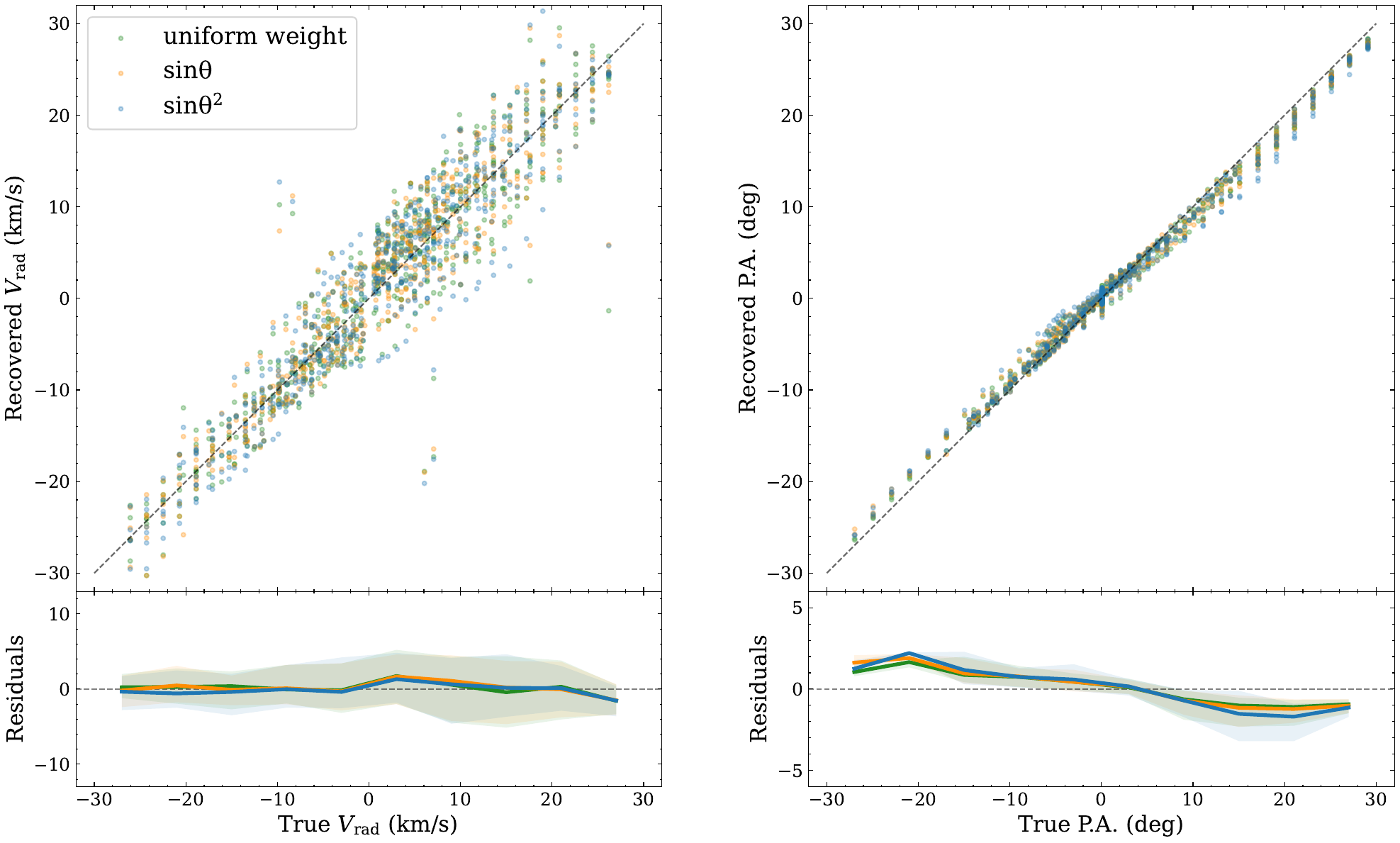}
    \caption{
    Comparison of $V_{\rm rad}$ (left) and position angle (PA; right) recovered with different weighting schemes.
    }
    \label{fig:weigting}
\end{figure*}

In our method, we adopt a $|\sin\theta \,|$ weighting scheme when fitting the kinematics with \texttt{$^{\rm 3D}$Barolo}. In this section, we assess how different weighting prescriptions affect the recovery of radial motions.
We tested three alternatives in the second step of our method: uniform, $| \sin\theta \,|$, and $\sin^2\theta$, using the galaxy sample described in Section~\ref{subsec:di}. The uniform weighting assigns equal importance to all azimuthal angles, thereby fully exploiting the available information across the disc. The $|\sin\theta\,|$ scheme, adopted as our default, enhances the contribution from regions close to the minor axis, where the radial component is maximally projected along the line of sight. The $\sin^2\theta$ weighting further amplifies this emphasis on the minor-axis regions.

The results are shown in Figure~\ref{fig:weigting}. For $V_{\rm rad}$, all three schemes recover the input values well across the full velocity range, with residuals typically within $\pm 2\ \mathrm{km\,s^{-1}}$. The uniform and $|\sin\theta\,|$ weightings show a slightly smaller scatter than the $\sin^2\theta$ case at negative radial velocities. Conversely, at positive radial velocities, the $\sin^2\theta$ scheme yields a marginally reduced scatter compared to the other two. 

For PA, all three weightings exhibit a mild systematic trend, although the absolute level of the residuals remains low. The largest scatter is observed for the $\sin^2\theta$ case, although the differences among the schemes are modest. 

Overall, these results indicate that the choice of weighting in the third and fourth steps has only a limited impact on the recovery of radial motions. Thus, the weighting scheme plays a much weaker role than the effect of simultaneously fitting radial motions together with the geometrical parameters.

\section{Inclination effect in comparison with previous approach}
\label{appendix:inc_di}
\begin{figure*}[htbp]
    \sidecaption
    \includegraphics[width=12cm]{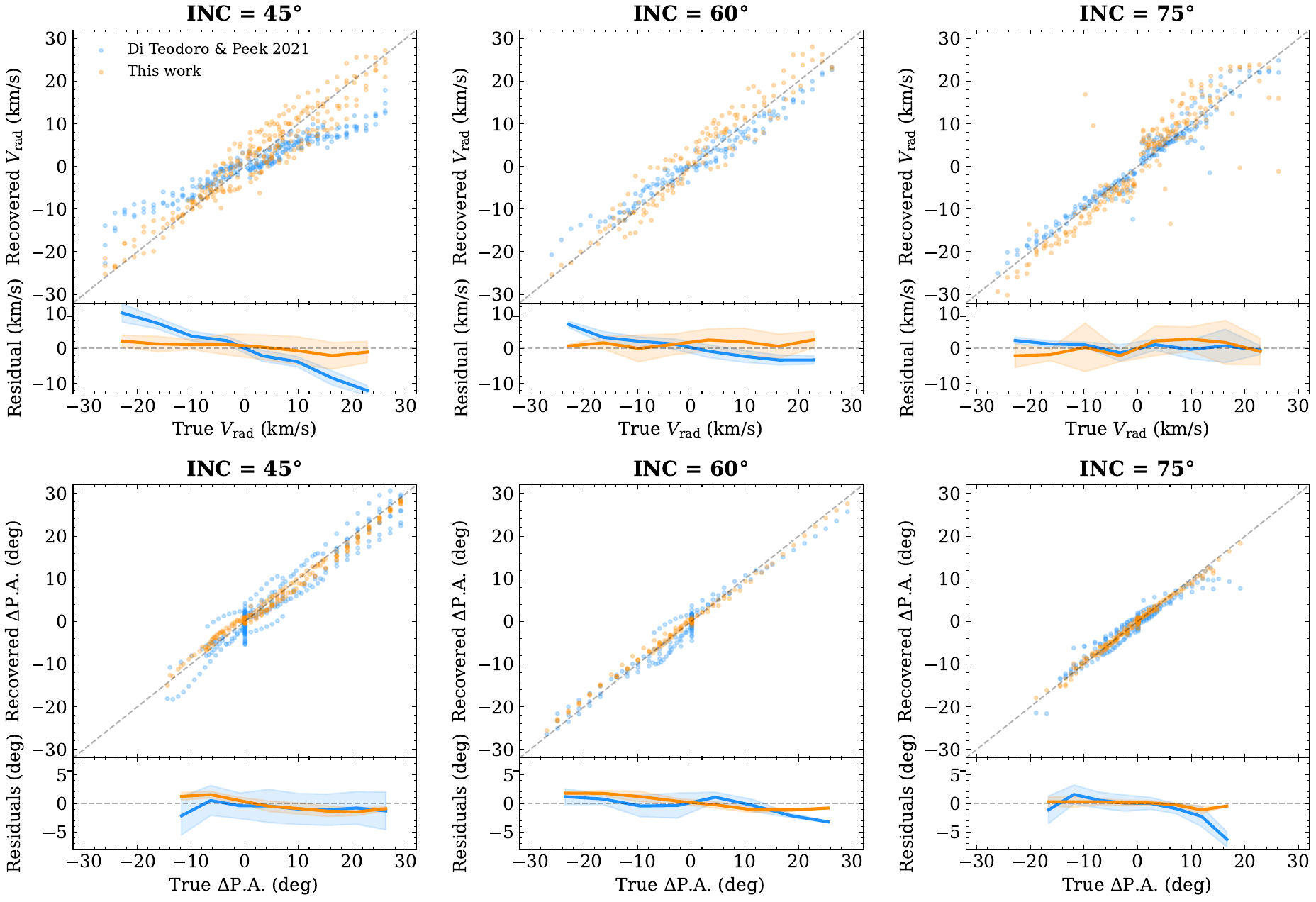}
    \caption{
    Same as Fig.~\ref{fig:comparisonwithenrico}, but shown separately for fixed inclinations of $45^\circ$, $60^\circ$, and $75^\circ$. The top row presents the comparison between true and recovered $V_{\rm rad}$, while the bottom row shows the corresponding results for the position angle.
    }
    \label{fig:inc_di}
\end{figure*}
In Fig.~\ref{fig:inc_di}, we present the comparison between our method and the approach of \citet{2021ApJ...923..220D} for three fixed inclinations ($i = 45^\circ, 60^\circ, 75^\circ$).
At $i = 45^\circ$, the method of \cite{2021ApJ...923..220D} exhibits a clear systematic trend: the residuals (Recovered $-$ True) decrease monotonically with increasing $V_{\rm rad}$, indicating a systematic underestimation of the radial velocity amplitude. In contrast, our method substantially reduces this bias, with residuals remaining close to zero over the full velocity range.
At $i = 60^\circ$, both methods perform significantly better, although a mild residual trend is still visible in the \cite{2021ApJ...923..220D} results.
At $i = 75^\circ$, the recovery improves for both approaches, with residuals tightly clustered around zero and no evident large-scale bias.

For the position angle, no strong systematic offset is observed in either method. However, at $i = 45^\circ$, the results of \cite{2021ApJ...923..220D} show noticeably greater scatter. At higher inclinations ($60^\circ$ and $75^\circ$), both methods converge toward nearly unbiased recovery, with reduced scatter.

The increased difficulty in disentangling $V_{\rm rad}$ and PA at low inclination has also been discussed by \citet{2023ApJ...944..143W}. At low $i$, the distortion of the velocity field induced by radial motions along the projected major axis becomes comparable to distortions along the minor axis. This geometric degeneracy closely resembles the kinematic signature of a warp. Consequently, fitting the geometric parameters first can partially absorb the radial flow signal into variations of the position angle, effectively masking part of the true radial motion. By fitting radial flows and geometric distortions simultaneously rather than sequentially, our method reduces this degeneracy and yields a significantly improved recovery in low-inclination systems.

\end{appendix}

\end{document}